\documentclass[fleqn,usenatbib]{mnras}

\usepackage{newtxtext,newtxmath}

\usepackage[T1]{fontenc}
\usepackage{ae,aecompl}

\usepackage{amsmath,amstext}
\usepackage{longtable}
\usepackage{comment}

\usepackage{graphicx}   
\usepackage{amsmath}    
\usepackage{amssymb}    
\usepackage{pdflscape}

\newcommand{\pdf}{${\rm PDF}(s)$} 
\newcommand{\pdfh}{${\rm PDF}(s_h)$} 
\newcommand{\jj}{$s$}
\newcommand{\jjh}{$s_h$}

\title[\pdf\/ -- morphology relation]{Stellar angular momentum distribution linked to galaxy morphology}

\author[Sarah M. Sweet et al.]{
Sarah M. Sweet,$^{1,2,3}$\thanks{E-mail: sarah@sarahsweet.com.au}
Karl Glazebrook,$^{2}$
Danail Obreschkow,$^{4}$
Deanne B. Fisher,$^{2}$
\newauthor
Andreas Burkert,$^{5,6}$
Claudia D. P. Lagos,$^{3,4}$
Juan M. Espejo Salcedo$^{2}$
\\
$^{1}$School of Mathematics and Physics, University of Queensland, Brisbane, QLD 4072, Australia\\
$^{2}$Centre for Astrophysics and Supercomputing, Swinburne University of Technology, PO Box 218, Hawthorn, VIC 3122, Australia\\
$^{3}$ARC Centre of Excellence for All Sky Astrophysics in 3 Dimensions (ASTRO 3D)\\
$^{4}$International Centre for Radio Astronomy Research, University of Western Australia, 7 Fairway, Crawley, WA 6009, Australia\\
$^{5}$Ludwig-Maximilians University Munich, University Observatory, Scheinerstr. 1, D-81679 M{\"u}nchen, Germany\\
$^{6}$Max-Planck Institute for Extraterrestrial Physics, Giessenbachstr. 1, D-85748 Garching, Germany
}

\date{Accepted XXX. Received YYY; in original form ZZZ}

\pubyear{2019}

\begin{document}
\label{firstpage}
\pagerange{\pageref{firstpage}--\pageref{lastpage}}
\maketitle

\begin{abstract}
We study the spatially-resolved stellar specific angular momentum $j_*$ in a high-quality sample of 24 CALIFA galaxies covering a broad range of visual morphology, accounting for stellar velocity and velocity dispersion. The shape of the spaxel-wise probability density function of normalised $s=j_*/j_{*mean}$, \pdf\/, deviates significantly from the near-universal initial distribution expected of baryons in a dark matter halo and can be explained by the expected baryonic effects in galaxy formation that remove and redistribute angular momentum. Further we find that the observed shape of the \pdf\/ correlates significantly with photometric morphology, where late-type galaxies have a \pdf\/ that is similar to a normal distribution, whereas early types have a strongly-skewed  \pdf\/ resulting from an excess of low-angular momentum material. Galaxies that are known to host pseudobulges (bulge S{\'e}rsic index $n_b <2.2$) tend to have less skewed bulge \pdf\/, with skewness $(b_{1rb})\lesssim0.8$. 
The \pdf\/ encodes both kinematic and photometric information and appears to be a robust tracer of morphology. Its use is motivated by the desire to move away from traditional component-based classifications which are subject to observer bias, to classification on a galaxy's fundamental (stellar mass, angular momentum) properties. In future, \pdf\/ may also be useful as a kinematic decomposition tool. 
\end{abstract}

\begin{keywords}
galaxies: bulges ---
galaxies: evolution ---
galaxies: fundamental parameters ---
galaxies: kinematics and dynamics ---
galaxies: spiral ---
galaxies: elliptical and lenticular, cD
\end{keywords}



\section{Introduction}
During the protogalactic growth of structure, large-scale tidal torques in the cosmic density field spin up the growing dark matter haloes \citep{Peebles1969,Doroshkevich1970,White1984}. 
The angular momentum of the haloes provides a seed for the angular momentum of the galaxies that form at the centres of the haloes \citep{White+Rees1978,Zavala+2016}.
Although the precise connection between the angular momenta of the galaxies and their haloes remains a topic of active research \citep[e.g.][]{Jiang+2019,Posti+2019}, 
it is well established that the angular momentum of galaxies plays a fundamental role for their global appearance, such as their size and density \citep{Mo+1998}. 
The co-evolution of a galaxy’s mass and angular momentum thus influences local \citep{Toomre1964} and global \citep{Ostriker+Peebles1973} stability modes, which is a likely reason for angular momentum to correlate with Hubble morphology \citep{Fall1983,RF12,OG14}, disk thickness and colour \citep{Hernandez+2006}, star formation efficiency \citep{Moster+2010} and cold gas fraction \citep{Huang+2012,Obreschkow+2016}.

The relationship between stellar mass $M_*$, stellar angular momentum $J_*$ and morphology has been explored since \citet{Fall1983}. Since the angular momentum of a mechanical system is proportional to its mass (if keeping lengths and velocities fixed), it is common to remove this obvious mass-dependence by introducing the specific angular momentum $j_* = J_*/M_*$. Any residual dependence between $j_*$ and $M_*$ then requires a non-trivial explanation. 
Empirically, more massive galaxies have higher angular momentum, with $j_* = qM_*^\alpha$. The morphology-dependent scale factor $q$ indicates a lower angular momentum for bulge-dominated galaxies \citep{RF12,Posti+2018,FR18}, and the roughly constant exponent $\alpha \sim 2/3$ is consistent with the corresponding exponent for haloes in a scale-free cold dark matter (CDM) universe \citep{Mo+1998}. However, at fixed bulge-to-total light ratio $\beta$, the data for spiral galaxies is consistent with  $\alpha = 1$ \citep{OG14,Cortese+2016,Sweet+2018}. In particular, galaxies that host rotation-supported pseudobulges follow along a tight track in $\beta$--$j_*/M_*$, consistent with $\alpha = 1$, whereas galaxies that host dispersion-supported classical bulges have larger bulges for a given $j_*/M_*$ \citep{Sweet+2018}, consistent with $\alpha = 2/3$ \citep{FR18}. Conversely, turbulent, clumpy galaxies thought to be the local analogues of high-redshift galaxies have smaller bulges at fixed $j_*/M_*$ than described by the `pseudobulge track', possibly indicating that their clumps may form future pseudobulges as they lose angular momentum and migrate to the centres of these galaxies \citep{Sweet+2019}.


Early toy models of disks \citep[e.g.][]{Fall+1980} assumed that the specific angular momentum of the baryons in the disk $j_{\rm b}$ follows that of the DM halo, $j_{\rm h}$, since the DM halo spinning up in the early universe is subject to the same tidal torques as the baryons it contains. This assumption has provided a working empirical description \citep[e.g. since][]{Fall1983}. 
However, the connection is not well understood: while it is reasonable that the baryons in the halo intially share the same $j$ as the DM, it does not necessarily follow that the baryons that collapse into a disk also share the same $j$. Since only a fraction of the baryons collapse into the disk, one naively expects that it is those with low $j$ that collapse first { \citep{Navarro+1997,Dutton+2012}}. This is indeed what early simulations showed, where disks in smoothed-particle hydrodynamics simulations historically had an excess of low-angular momentum material, manifested as larger bulges and thicker disks than observed \citep{Governato+2010,Agertz+2011}. This ``angular momentum catastrophe'' is somewhat ameliorated by including appropriate feedback mechanisms to remove low-angular momentum material \citep[e.g.][]{Genel+2015}. However, this fine tuning does not explain the broad correspondence between $j_{\rm b}$ and $j_{\rm h}$, wherein the ratio $j_{\rm b}/j_{\rm h}$ must vary with $(M_{\rm b}/M_{\rm h})^{2/3}$ to give $j_{\rm b} \sim M_{\rm b}^{2/3}$ \citep{Posti2018MN}.

While the mean $j_{\rm b}$ and $j_{\rm h}$ agree within a factor of two, 
the internal angular momentum distribution of the baryons is 
substantially affected by the physical processes that rearrange and remove angular momentum and stellar mass in the disk. For example,
stellar and AGN feedback remove low-angular momentum material, and tidal stripping preferentially removes high-angular momentum material { \citep{Governato+2010,vandenBosch2001}}. Viscosity, resulting from bars and star-forming clumps, transports angular momentum in the disk and impacts the angular momentum distribution \citep{vandenBosch+2001,Elmegreen+2013}. The angular momentum distribution is also affected by galaxy mergers. \citet{Lagos+2018} used the EAGLE cosmological hydrodynamical simulations 
\citep{Schaye+2015} to quantify the effect. They found that gas-poor (dry) mergers significantly redistribute the stellar $j_*$ in a way that the internal parts become largely $j_*$-deficient, while gas-rich (wet) mergers can have the opposite effect, of increasing $j_*$ in the central regions.  The angular momentum distribution thus promises a stronger tracer of a galaxy's evolutionary history than mean kinematic properties such as $j_*$ or global spin parameter $\lambda$. 

The angular momentum distribution is thought to be near-universal for DM haloes owing to the balance between cosmic torques imparting angular momentum and the mass profile of the material to which it is imparted \citep{vandenBosch+2001,Bullock+2001,Liao+2017}. This universality has provided clear predictions for the distribution of angular momentum in haloes and inspired much effort to link to observed angular momentum, using various methods to quantify angular momentum and its distribution. A simple quantification is found in specific angular momentum $j = J/M = \int dM \vert {\bf r} \times {\bf v}\vert /M$, which is defined and measurable both for stars in the disk ($j_*$) and for the halo ($j_{\rm h}$). 

The distribution of $j_*$ has been studied as early as \citet{Crampin+1964}, who found the mass distribution of $j_*$ for eight Sb and Sc spiral galaxies to be consistent with the mass distribution of a uniformly-rotating spheroid. \citet{Fall+1980} then explored the formation of model disk galaxies embedded in dark matter haloes and also found the angular momentum of the disk to be consistent with that of an isothermal sphere. However, it was later shown that the distribution of $j_*$ in the baryons ought to be further affected by nonlinear torques \citep{Barnes+1987}, so the link between the the angular momentum distributions for baryonic disk and DM halo is likely more complicated than first appeared.  
\citet{Bullock+2001} used a shape parameter $\mu$ to parametrize the cumulative mass distribution of $j$ in haloes, and found a narrow distribution around mean $\rm{log}(\mu-1) = -0.6\pm0.4$, having little or no correlation with mass or morphology; $\mu$ is larger for the (gas) disk than for the DM halo \citep{Chen+2003}. \citet{vandenBosch+2001} showed for the same DM haloes in \citet{Bullock+2001} that the mean probability density function (PDF) of normalised \jjh\/$=j/j_{mean}$ is described by a distribution that peaks at $j=0$ with a tail to high $j$, with the exact functional form depending on $\mu$. Later study of the misalignment between gas and DM demonstrated a component of negative angular momentum material contributing to a bulge and leaving behind an exponential density distribution \citep{vandenBosch+2002}. Exponential disks have also been shown to form ubiquitously from stellar scattering off of star-forming clumps \citep{Elmegreen+2013}, and radial migration \citep{Herpich+2017}. 
\citet{vandenBosch+2001} were among the first to remark that rotating disk galaxies (in their case, dwarfs) are deficient both in high- and low-$j_*$ material when compared with their parent DM halo. 
In order to describe both centrally-peaked halo PDFs and exponential disk PDFs with a single functional form, \citet{SS05} parametrized \jj\/ by a Gamma distribution function. They found a mean shape parameter\footnote{Referred to here with subscript so as not to be confused with slope of the $M_*-j_*$ relation.} $\alpha_\Gamma = 0.83$ for haloes and $\alpha_\Gamma = 1.41$ for model exponential disks embedded in a NFW halo \citep{Sharma+2012}. However, since the Gamma function diverges, such fits are unstable near the critical value at $\alpha_\Gamma = 1$ and this parametrization is consequently not usefully robust for most observations.

Regardless of the parametrization, the \pdf\ of the stars encodes much physical information about the evolution of the baryonic material after becoming decoupled from the DM halo. Outflowing material due to AGN- or stellar feedback carries away inner, dispersion-dominated gas with low \jj\/, leading to lower angular momentum of the stars that later form from that gas {\citep{Dutton2009,Brook+2011,Sharma+2012}}, with low-mass galaxies losing the bulk of such material but high-mass galaxies potentially causing a galactic fountain, where the outflowing material falls back onto the disk and is redistributed \citep{Brook+2012}. Conversely, tidal stripping by interactions with the intergalactic medium removes outer, rapidly-rotating, high-\jj\/ material \citep{vandenBosch+2001}. Angular momentum is also redistributed by star-forming clumps, spiral arms, bars, mergers and viscosity \citep{Mo+1998}. Biased collapse of preferentially low-angular momentum material  \citep{Posti2018MN} is another mechanism for differences between the stellar and halo angular momentum distribution. Given the near-universality of halo \pdfh\/ and the fact that the distributions of the halo and baryonic $j$ are initially linked due to common tidal torques, observed differences between the distributions trace the physical processes that affect the baryons and produce the observed morphologies \citep{vandenBosch+2001}. 
In addition, the observable stellar \pdf\/ combines both photometric (surface density) and kinematic information in a spatially-resolved manner and may be useful as a tracer of morphology and as a bulge-disk decomposition tool. 

The motivation for this paper is that there is no sample of nearby galaxies spanning morphological types E0 to Sd for which \pdf\/ has been measured, yet quantitative measures of \pdf\/ can provide a new, more physically-motivated way to compare simulations and data, where additional (kinematic) information is added to visual morphology. Accordingly, in this paper we present stellar specific angular momentum distributions for 24 nearby galaxies with high-quality observations in the Calar Alto Legacy Integral Field Area survey \citep[CALIFA;][]{Sanchez+2012,Husemann+2013,Walcher+2014,Sanchez+2016}, and we develop and provide statistical parametrizations of the shapes that can be easily compared with those from simulations. Section~\ref{sec:sample} describes the data and the observations. Our methods for computing spatially-resolved specific angular momentum are described in Section~\ref{sec:methods}. Parametrizations of \pdf\/ are presented in Section~\ref{sec:results}. In Section~\ref{sec:analysis} we investigate the \pdf\/-morphology relation. Implications for disk evolution and the utility of \pdf\/ as a kinematic tracer of morphology and decomposition tool are discussed in Section~\ref{sec:discussion}. Section~\ref{sec:conclusions} concludes the paper. 

We assume a cosmology where $H_0 = 70 \rm{km s}^{-1} \rm{Mpc}^{-1}$, $\Omega_M = 0.27$ and $\Omega_\Lambda = 0.73$, and quote comoving coordinates throughout. 

\section{Sample \& Observations}
\label{sec:sample}

The CALIFA survey presented stellar kinematic and surface density maps for 300 nearby galaxies in \citet{Falcon-Barroso+2017}. The maps are based on the V1200 grating, with wavelength range 3750--4550\AA\/, instrumental resolution of approximately 72 km s$^{-1}$, and 1$\arcsec$ spaxels. The spatial resolution is limited in the galaxy centres by the 2.7$\arcsec$-wide fibres and at the outskirts by the Voronoi binning \citep{Cappellari+2003} used to increase signal-to-noise per bin above 20. The authors employed penalised pixel fitting pPXF \citep{Cappellari+2004}, using spectral templates from \citet{Valdes+2004}, to obtain Gaussian line-of-sight velocities and velocity dispersions. 
In \citet{Sweet+2018} we inspected the integrated angular momentum $j_*(<r)$ for CALIFA galaxies observed to the highest multiples of effective radius $r_e$, and found that $j_*(<r)$  for those galaxies typically reaches $>0.99j_*$ at a radius $r \sim 3r_e$. We thus selected galaxies observed to at least $3r_e$, which resulted in a sample of 50 galaxies. For this work we focus on galaxies without bars in order to simplify our analysis of the relation between \pdf\/ and morphology. After discarding barred galaxies we are left with 26 regular local galaxies with high-quality kinematics. We also discard two galaxies with skewness $b_1$ or $b_{1rb}$ (as defined in Section~\ref{sec:methods}) more than 15$\sigma$ from the sample mean. 
Bulge-to-total light ratios $\beta$ are taken from bulge-disk decompositions given in \citet{Mendez-Abreu+2017}, who use a Levenberg-Marquardt algorithm to fit components to the 2D surface brightness distribution of Sloan Digital Sky Survey \citep[SDSS,][]{Abazajian+2009} images. We use the $r$-band decompositions in this work. We obtain Hubble types $T$ using the methods of \citet{fisher2008}, namely combining high-resolution imaging from HST with wide-field ground-based imaging, and accounting for a different mass-to-light ratio in the bulge versus the disk. These $T$ are robust to within an uncertainty of $\Delta T = \pm 0.5$. The $M_*-j_*$ relation is shown in  Figure~\ref{fig:m_j}. SDSS postage stamps for each galaxy in our sample are shown in Appendix~\ref{app:thumbs}.
The final sample of 24 galaxies spans bulge-to-total ratios $0.02\leq \beta \leq 0.67$ and Hubble types from Sd to E/S0.

\begin{figure}
    \centering
    \includegraphics[width=1.1\linewidth]{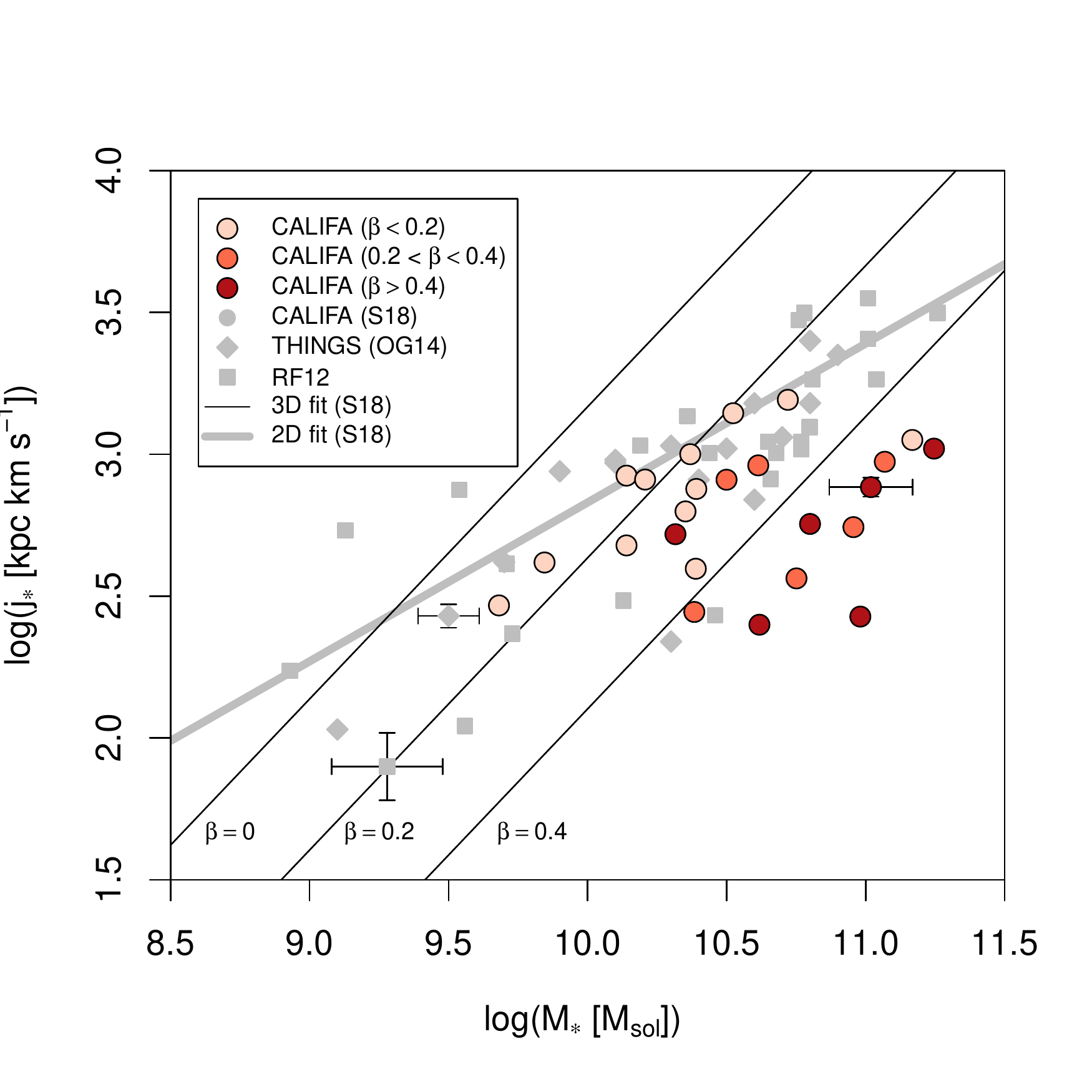}
\caption{The $M_*-j_*$ plane. 
CALIFA galaxies in this work are shown as large filled circles, assigned discrete colour shades according to bins of $\beta ~\epsilon~ [0, 0.2, 0.4, 1)$. 
Samples from the literature are shown as grey solid symbols.
A typical error bar is shown for each of the samples.
Best-fitting lines are from \citet{Sweet+2018}. Solid lines of constant $\beta$ represent a trivariate fit in $M_*-j_*-\beta$ space, while the thick grey line represents the 2D fit. 
The galaxies in this work occupy a similar parameter space to the parent sample of 50 galaxies presented in \citet{Sweet+2018}.}
    \label{fig:m_j}
\end{figure}


\section{Methods} 
\label{sec:methods}

Specific angular momentum is measured in a spatially-resolved manner as follows. 
First, in every spaxel $i$ within the observed map of $n$ spaxels, we calculate the $z-$component (along the rotation axis) of the mean stellar specific angular momentum,
\begin{equation}
j_{*,i} = ({\bf r}_i \times {\bf v}_{i})_z,
\label{eq:ji}
\end{equation}
where ${\bf v}_{i}$ is the stellar velocity in that spaxel and ${\bf r}_i$ is the radius of that spaxel from the galaxy centre. Since only the line-of-sight component of ${\bf v}_{i}$ and the transverse components of ${\bf r}_i$ can be measured, we recover the remaining vector components by assuming that all galaxies are flat circular disks of radius-independent inclination and position angle derived from a fit to the CALIFA $B$-band surface brightness maps.

Note that each spaxel $i$ contains a large number of unresolved particles (stars) on different orbits, as evidenced by the measurable line-of-sight velocity dispersion ${\sigma}_{i}$. The velocity distribution is assumed to be isotropic and normal. In this case, the distribution of specific angular momenta in each spaxel is also normally distributed, with a mean given by Eq.~\ref{eq:ji} and standard deviation $\vert {\bf r}_i \vert \sigma_i$. Formally, each spaxel $i$ gives rise to a normal $j_*$-distribution expressed as
\begin{equation}
    {\rm PDF}_i(j_{*}) = \mathcal{N} (j_{*}|\mu=j_{*,i},\sigma^2=|{\bf r}_i|^2\sigma_{i}^2).
\end{equation}
The total ${\rm PDF}(j_{*})$ of the full stellar disk can then be computed as the mass-weighted average of ${\rm PDF}_i(j_{*})$,
\begin{equation}
    {\rm PDF}(j_{*}) = \frac{\sum_{i=1}^n\,m_{*,i}\,{\rm PDF}_i(j_{*})}{M_*},
\label{eq:pdf}
\end{equation}
where $m_{*,i}$ is the stellar mass in the $i$-th spaxel and $M_{*,i}\equiv\sum_{i=1}^n m_{*,i}$ is the total stellar mass. Upon assuming a constant mass-to-light ratio ($M/L$) across the galaxy, the actual value of $M/L$ is irrelevant, since it appears equally in the numerator and denominator of Eq.~\ref{eq:pdf} and hence cancels out.

It is convenient to introduce the \textit{normalised} specific angular momentum $s=j_*/j_{*\rm mean}$, where
\begin{equation}
    j_{*\rm mean} = J_*/M_* 
    = \frac{\sum_{i=1}^n m_{*,i}\left({\bf r}_i\times{\bf v}_i\right)_z}{\sum_i^n m_{*,i}}=M_*^{-1}\sum_{i=1}^n m_{*,i}~j_{*,i}.
\end{equation}
From here on, ${\rm PDF}(s)$ is defined as Eq.~\ref{eq:pdf}, where the argument is $j_*=s\,j_{*\rm mean}$.

The probability density in a finite interval $[s,s+\Delta s]$ can be computed as
\begin{equation}
\begin{split}
    & P(s,s+\Delta s) = \int_{s}^{s+\Delta s} {\rm PDF}(x)\,{\rm d}x \\
    & = \frac{1}{2M_*}\sum_{i=1}^n m_{*,i}\left[{\rm erf}\left(\frac{s+\Delta s-s_{i}}{\sqrt{2}\sigma_{s,i}}\right)-{\rm erf}\left(\frac{s-s_{i}}{\sqrt{2}\sigma_{s,i}}\right)\right],   
    \label{eq:ps}
\end{split}
\end{equation}
where $s_{i}=j_{*,i}/j_{*\rm mean}$ and $\sigma_{s,i}=\vert {\bf r}_i \vert \sigma_i/j_{*\rm mean}$. Discretised probability densities can thus be obtained as $P(s,s+\Delta s)/\Delta s$.
For plotting purposes, we will choose $\Delta s$ such that the number of histogram bins equals the number of seeing elements along the major axis of the galaxy, but all our statistical analyses rely on much smaller (numerically converged) bins.

In addition to the \emph{dispersion-broadened} \pdf\/ described above, we also derive a \emph{circular-only} \pdf\/, that is, setting the dispersion to zero. The motivation for this kind of \pdf\/ is to provide a benchmark comparison with samples that do not have dispersion measurements e.g. observations of gas kinematics-only, at low spectral resolution, and/or at high-redshift. Equation~\ref{eq:ps} is undefined when dispersion is zero, so the circular-only \pdf\/ is calculated by setting the dispersion to be small (in practise, $\sigma_{s,i} \sim {\bf v}_i/1e4 $); this is appropriate when the number of spaxels across the galaxy is correspondingly large.

We can analyse the shape of ${\rm PDF}(s)$, for instance by analysing the moments of this continuous distribution. The mean $\mu = E(s) \equiv 1$ by definition of $s$. The $n$-th moment is given by
\begin{equation}
    \mu_n = \int (s - \mu)^n {\rm PDF}(s) {\rm d}s.
    \label{eq:mun}
\end{equation}
We evaluate this integral numerically between $\pm A$, where $A = \rm{max}_i \{(j_{*,i}+3\vert {\bf r}_i \vert \sigma_i)/j_{*\rm mean}\}$. 
The standard deviation $\sigma = \sqrt{\mathrm{E}\left[(s-\mu)^{2}\right]}$ is the square root of the second central moment of the distribution.
The third standardised moment about the mean quantifies skewness\footnote{The notation $b_1$, $b_2$ refers to the sample statistic. We use this notation rather than population $\beta_1$, $\beta_2$ to avoid confusion with bulge-to-total ratio $\beta$ as commonly used in angular momentum studies.}:
\begin{equation}
b_1 = \frac{\mu_{3}}{\sigma^{3}}=\frac{\mathrm{E}\left[(s-\mu)^{3}\right]}{\sigma^{3}},
\end{equation}
Distributions that are symmetric about the mean have $b_1 = 0$. Higher $b_1 > 0$ indicates more probability density in the tail to the right of the mean (\emph{positively skewed}), and $b_1 < 0$ indicates more probability density in the left tail (\emph{negatively skewed}).

The fourth standardised moment about the mean quantifies kurtosis:
\begin{equation}
b_2 = \frac{\mu_{4}}{\sigma^{4}}=\frac{\mathrm{E}\left[(s-\mu)^{4}\right]}{\sigma^{4}}.
\label{eq:kurt}
\end{equation}
Kurtosis $b_2 = 3$ corresponds to a Gaussian-like (\emph{mesokurtic}) distribution. Higher $b_2 > 3$ indicates that there is more probability density in the peak and tails rather than the shoulders of the distribution compared to a normal distribution (\emph{leptokurtic}). Lower $b_2 < 3$ indicates that there is less probability density in the peak and tails (\emph{platykurtic}). 

Uncertainties are estimated by conducting a Monte-Carlo simulation where the noise term added to ${\bf v}_i$ is randomly drawn from a normal distribution whose width is the stellar circular velocity 1-$\sigma$ uncertainty map; the noise added to $\sigma_i$ is similarly drawn from the velocity dispersion 1-$\sigma$ uncertainty map.

\begin{figure*}
    \centering
    \includegraphics[width=1.1\linewidth]{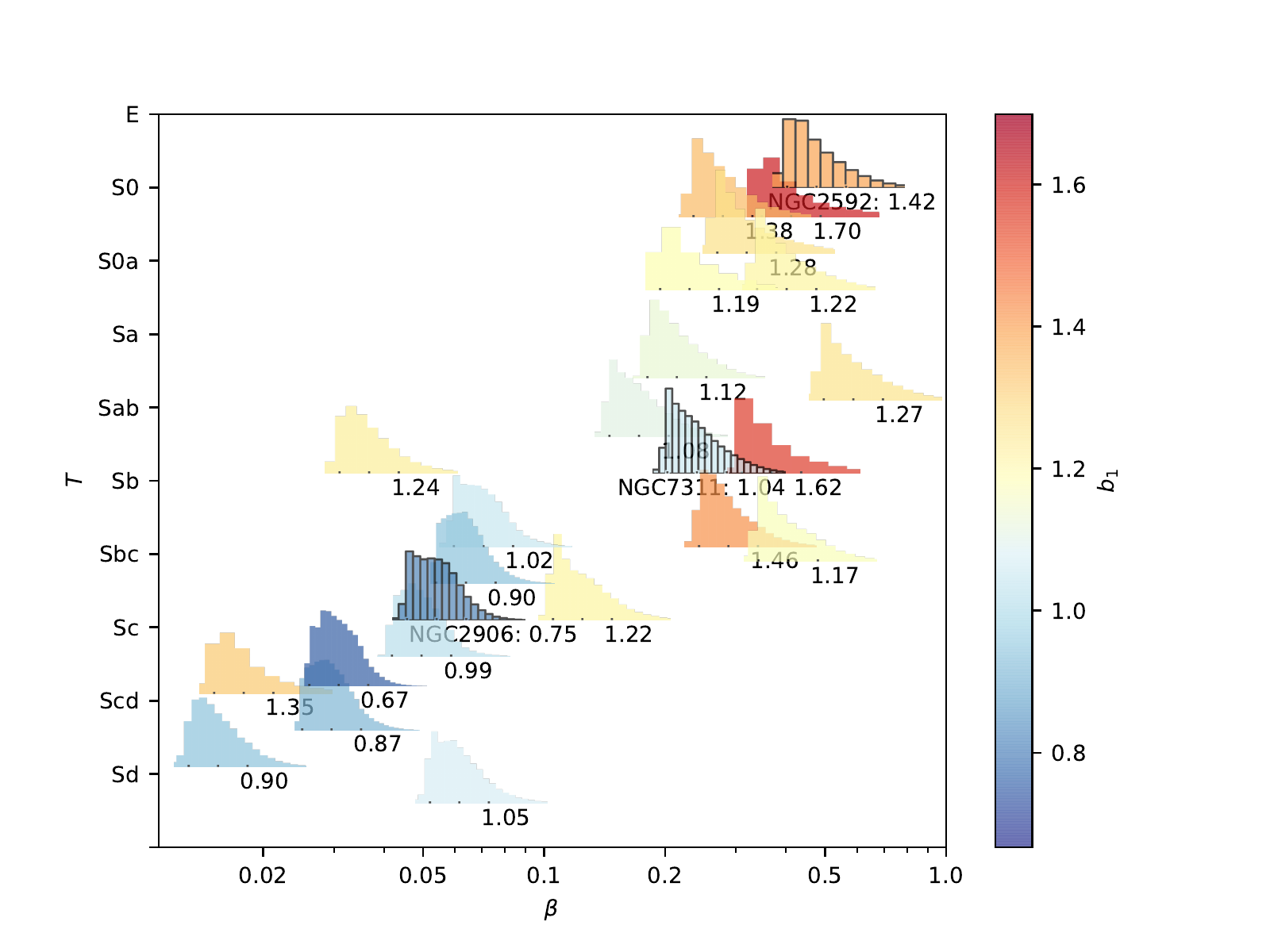}
    \caption{Probability density functions of normalised specific angular momentum (\pdf\/) for CALIFA galaxies, plotted in bulge-to-total light ratio $\beta$ - Hubble type $T$ space. An offset of up to $\pm 0.5 \Delta T$ is added in the $y$-direction  for clarity. Each \pdf\/ has ticks at \jj=0, 1, and 2, and is labelled with $b_1$. Highlighted with black outlines are three examples examined further throughout the paper: {\bf 1)} late-type spiral NGC 2906 with low $\beta = 0.06$ has a broad, symmetric \pdf\/ centred on \jj\/ = 1. {\bf 2)} early type NGC 2592 with high $\beta = 0.54$ has a strongly-skewed \pdf\/ which peaks nearer \jj\/ = 0. {\bf 3)} NGC 7311 with moderate $\beta = 0.27$ has a \pdf\/ which is intermediate between the two extremes. 
    The shape of \pdf\/ is quantified by skewness $b_1 = \mu_3 / \sigma^3$ as defined in the text. More normal, symmetric distributions ($b_1 \sim 0.8$) are coloured blue and strongly-skewed ($b_1 > 1.2$) distributions are coloured red.
    Galaxies with earlier types and larger $\beta$ tend to have skewed \pdf\/, while the later types with smaller $\beta$ have symmetric \pdf\/.
    }
    \label{fig:histograms}
\end{figure*}

Results and properties of the sample of CALIFA galaxies are given in Table~\ref{tab:results}. Plots of the ${\rm PDF}(s)$ are shown in Figure~\ref{fig:histograms}.

\begin{table*}
\centering
\caption{Properties of CALIFA galaxies.} 
\label{tab:results}
\begin{tabular}{lrrrrrrrrrrrrrrrrrrrrrrrr}
  \hline
Name & RA & Dec & $z$  & $M_*$ & $\Delta M_*$ & $\beta $ & $ \Delta \beta$ & $n_b$ & $\Delta n_b$ & $T$ \\ 
(1)  & (2) & (3) & (4) & (5) & (6) & (7) & (8) & (9) & (10) & (11)   \\
  \hline
IC1151 & 239.6347351 & 17.4414577 & 0.009 & 9.85 & 0.15 & 0.02 & 0.10 & 0.6 & 0.0 & Scd \\ 
  MCG-02-02-030 & 7.5304556 & -11.1136293 & 0.011 & 10.37 & 0.15 & 0.08 & 0.10 & 1.2 & 0.1 & Sb \\ 
  NGC0001 & 1.8160844 & 27.7080822 & 0.015 & 10.80 & 0.15 & 0.46 & 0.10 & 2.8 & 0.2 & Sb \\ 
  NGC2253 & 100.9243164 & 65.2063751 & 0.013 & 10.52 & 0.15 & 0.06 & 0.10 & 1.1 & 0.1 & Sbc \\ 
  NGC2592 & 126.7835464 & 25.9703159 & 0.008 & 10.62 & 0.15 & 0.54 & 0.10 & 3.3 & 0.1 & E?S0? \\ 
  NGC2639 & 130.9086151 & 50.2055397 & 0.012 & 11.17 & 0.15 & 0.20 & 0.10 & 2.0 & 0.1 & Sa \\ 
  NGC2906 & 143.0259094 & 8.4417686 & 0.008 & 10.39 & 0.15 & 0.06 & 0.10 & 1.4 & 0.0 & Sbc \\ 
  NGC3815 & 175.4137115 & 24.8004990 & 0.014 & 10.35 & 0.15 & 0.04 & 0.10 & 0.7 & 0.1 & Sab \\ 
  NGC4961 & 196.4481812 & 27.7341385 & 0.010 & 9.68 & 0.15 & 0.02 & 0.10 & 0.7 & 0.1 & Sc \\ 
  NGC5480 & 211.5899048 & 50.7250633 & 0.008 & 10.14 & 0.15 & 0.07 & 0.10 & 1.4 & 0.1 & Sd \\ 
  NGC5971 & 233.9037476 & 56.4616890 & 0.016 & 10.32 & 0.15 & 0.67 & 0.10 & 4.4 & 0.4 & Sa \\ 
  NGC5980 & 235.3768463 & 15.7876863 & 0.016 & 10.72 & 0.15 & 0.07 & 0.10 & 0.9 & 0.1 & Sbc \\ 
  NGC6063 & 241.8041382 & 7.9789910 & 0.011 & 10.14 & 0.15 & 0.03 & 0.10 & 3.6 & 0.3 & Sc \\ 
  NGC6427 & 265.9108276 & 25.4939384 & 0.013 & 10.75 & 0.15 & 0.36 & 0.10 & 2.3 & 0.3 & S0 \\ 
  NGC6762 & 286.4045410 & 63.9341087 & 0.011 & 10.38 & 0.15 & 0.32 & 0.10 & 3.5 & 0.3 & S0 \\ 
  NGC7311 & 338.5283203 & 5.5703239 & 0.015 & 11.07 & 0.15 & 0.27 & 0.10 & 1.6 & 0.1 & Sab \\ 
  NGC7623 & 350.1250305 & 8.3957691 & 0.013 & 10.98 & 0.15 & 0.47 & 0.10 & 2.0 & 0.2 & S0 \\ 
  NGC7653 & 351.2056580 & 15.2756014 & 0.014 & 10.50 & 0.15 & 0.33 & 0.10 & 3.3 & 0.2 & Sb \\ 
  NGC7671 & 351.8305664 & 12.4674091 & 0.014 & 10.96 & 0.15 & 0.26 & 0.10 & 2.1 & 0.1 & S0 \\ 
  NGC7683 & 352.2659302 & 11.4451685 & 0.012 & 11.02 & 0.15 & 0.46 & 0.10 & 2.4 & 0.1 & S0 \\ 
  NGC7716 & 354.1310425 & 0.2972720 & 0.009 & 10.39 & 0.15 & 0.14 & 0.10 & 1.8 & 0.2 & Sbc \\ 
  NGC7824 & 1.2759866 & 6.9201469 & 0.020 & 11.25 & 0.15 & 0.42 & 0.10 & 2.3 & 0.2 & Sab \\ 
  UGC00987 & 21.3810329 & 32.1362724 & 0.015 & 10.61 & 0.15 & 0.24 & 0.10 & 3.9 & 0.4 & Sa \\ 
  UGC09476 & 220.3834534 & 44.5127716 & 0.013 & 10.21 & 0.15 & 0.03 & 0.10 & 1.2 & 0.1 & Sc \\  
     \hline
\end{tabular}
\begin{tabular}{lrrrrrrrrrrrrrrr}
  \hline
Name  & $j_* $ & $ \Delta j_*$ & $P_{\beta}$ & $b_1$ & $\Delta b_1$ & $b_2$ & $\Delta b_2$ & $b_{1c}$ & $\Delta b_{1c}$ & $b_{1rb}$ & $\Delta b_{1rb}$ & $r_b$\\
(1) & (12) & (13) & (14) & (15) & (16) & (17) & (18) & (19) & (20) & (21) & (22) & (23) \\ 
  \hline
IC1151 & 416 & 57 & 0.48 & 0.900 & 0.020 & 4.906 & 0.117 & 0.637 & 0.014 & 0.223 & 0.141 & 7.52\\ 
  MCG-02-02-030 & 1001 & 126 & 0.47 & 1.017 & 0.013 & 4.783 & 0.076 & 0.677 & 0.007 & 0.377 & 0.023 & 6.32 \\ 
  NGC0001 & 568 & 41 & 0.53 & 1.169 & 0.015 & 4.980 & 0.091 & 1.006 & 0.008 & 0.908 & 0.025 & 2.53 \\ 
  NGC2253 & 1396 & 205 & 0.51 & 0.992 & 0.014 & 5.259 & 0.136 & 0.716 & 0.007 & 0.592 & 0.018 & 5.61\\ 
  NGC2592 & 251 & 10 & 0.50 & 1.423 & 0.010 & 6.302 & 0.110 & 1.241 & 0.009 & 0.736 & 0.009 & 2.20\\ 
  NGC2639 & 1123 & 106 & 0.45 & 1.082 & 0.005 & 4.507 & 0.028 & 0.564 & 0.003 & 0.619 & 0.009 & 4.47\\ 
  NGC2906 & 755 & 110 & 0.46 & 0.746 & 0.009 & 3.632 & 0.038 & 0.575 & 0.005 & 0.480 & 0.020 & 5.69\\ 
  NGC3815 & 629 & 70 &   & 1.241 & 0.018 & 7.018 & 0.136 & 0.961 & 0.021 & 0.381 & 0.022 & 4.88\\ 
  NGC4961 & 293 & 30 &   & 1.349 & 0.021 & 8.787 & 0.163 & 1.313 & 0.010 & 0.302 & 0.069 & 4.45\\ 
  NGC5480 & 477 & 70 & 0.45 & 1.049 & 0.017 & 5.256 & 0.103 & 0.868 & 0.009 & 0.498 & 0.062 & 6.12\\ 
  NGC5971 & 523 & 65 &   & 1.273 & 0.023 & 6.025 & 0.175 & 0.919 & 0.013 & 1.049 & 0.032 & 2.79\\ 
  NGC5980 & 1556 & 193 & 0.38 & 0.895 & 0.016 & 4.428 & 0.090 & 0.562 & 0.007 & 0.578 & 0.031 & 5.83\\ 
  NGC6063 & 841 & 116 & 0.53 & 0.668 & 0.015 & 3.473 & 0.057 & 0.277 & 0.005 & 0.457 & 0.125 & 7.58\\ 
  NGC6427 & 365 & 15 & 0.53 & 1.284 & 0.008 & 4.702 & 0.042 & 1.316 & 0.004 & 0.704 & 0.011 & 1.93\\ 
  NGC6762 & 278 & 20 &   & 1.380 & 0.028 & 5.970 & 0.208 & 1.188 & 0.012 & 0.748 & 0.024 & 2.72\\ 
  NGC7311 & 941 & 61 & 0.47 & 1.040 & 0.008 & 4.057 & 0.033 & 0.667 & 0.004 & 0.745 & 0.014 & 2.57\\ 
  NGC7623 & 268 & 11 &   & 1.699 & 0.010 & 7.088 & 0.068 & 1.920 & 0.008 & 0.431 & 0.008 & 2.05\\ 
  NGC7653 & 813 & 89 & 0.48 & 1.459 & 0.017 & 6.863 & 0.133 & 1.259 & 0.007 & 0.771 & 0.024 & 3.87\\ 
  NGC7671 & 554 & 23 &   & 1.190 & 0.008 & 4.886 & 0.041 & 0.908 & 0.006 & 0.415 & 0.009 & 2.33\\ 
  NGC7683 & 766 & 61 & 0.52 & 1.223 & 0.008 & 5.082 & 0.051 & 1.113 & 0.006 & 0.764 & 0.015 & 2.93\\ 
  NGC7716 & 395 & 23 & 0.45 & 1.222 & 0.011 & 5.209 & 0.067 & 0.788 & 0.005 & 0.532 & 0.010 & 2.80\\ 
  NGC7824 & 1047 & 40 &   & 1.624 & 0.041 & 9.885 & 1.014 & 1.531 & 0.013 & 0.782 & 0.025 & 2.14\\ 
  UGC00987 & 914 & 86 & 0.49 & 1.118 & 0.022 & 4.602 & 0.226 & 0.838 & 0.009 & 0.580 & 0.025 & 2.92\\ 
  UGC09476 & 815 & 116 & 0.46 & 0.866 & 0.016 & 4.229 & 0.088 & 0.500 & 0.006 & 0.578 & 0.146 & 6.28\\ 
     \hline
\end{tabular}\\
Columns: 
(1) galaxy identifier;  (2) right ascension (J2000) [deg];
 (3) declination (J2000) [deg];
 (4) redshift;
 (5) base 10 logarithm of stellar mass [log($M_\odot$)]; (6) measurement uncertainty in $M_*$ [dex]; (7) $r$-band bulge-to-total light ratio; (8) measurement uncertainty in $\beta$; 
 (9) bulge S{\'e}rsic index; (10) measurement uncertainty in $n_b$; (11) Hubble type;
(12) stellar specific angular momentum [kpc km s$^{-1}$]; (13) measurement uncertainty in $j_*$ [kpc km s$^{-1}$];
 (14) area under `bulge' curve from normal + log-normal fit (* denotes fit failed); (15)
 skewness of \pdf\/; (16) measurement uncertainty in $b_1$; (17)
 kurtosis of \pdf\/; (18) measurement uncertainty in $b_2$; (19) skewness of \pdf\/ without including velocity dispersion; (20) measurement uncertainty in $b_{1c}$; (21) skewness of radius-selected bulge \pdf\/; (22) measurement uncertainty in $b_{1rb}$; (23) radius of `bulge' component [\arcsec].
 (7-10) from CALIFA; (11-23) this work.
\end{table*} %

\section{Results} 
\label{sec:results}

In this Section we describe the probability density function of normalised specific angular momentum \pdf\/ and investigate methods for quantifying its shape, demonstrating that parametrization is not the best method for describing \pdf\/.

\subsection{Shape of \pdf\/}

Figure~\ref{fig:histograms} illustrates the range of \pdf\/ that characterises local CALIFA galaxies. For galaxies with earlier types and larger bulge-to-total ratio $\beta$, the \pdf\/ visually tends to have a sharper peak near \jj\/ = 0 and exponential decline, corresponding to a positively-skewed \pdf\/ with $b_1 > 1.2$.
On the other hand, the later type galaxies with smaller $\beta$ visually have a broader, more normal distribution and peak nearer \jj\/ = 1, have more symmetric \pdf\/ with $b_1 \sim 0.8$. 

\begin{figure*} 
\begin{center}
 \includegraphics[width=0.45\linewidth]{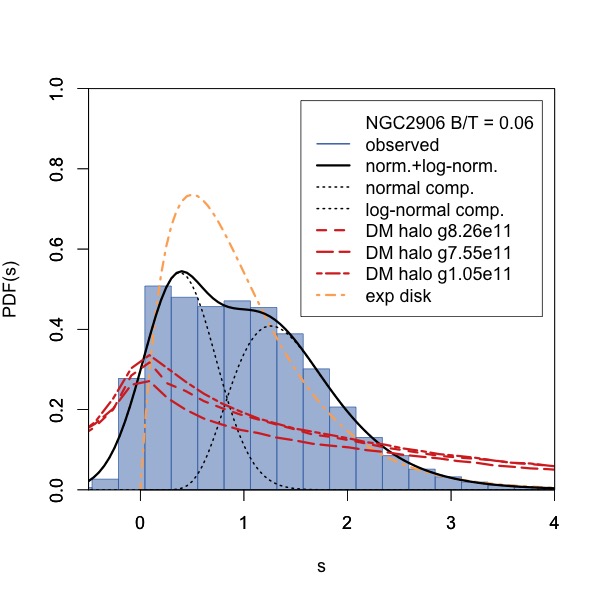}
 \includegraphics[width=0.45\linewidth]{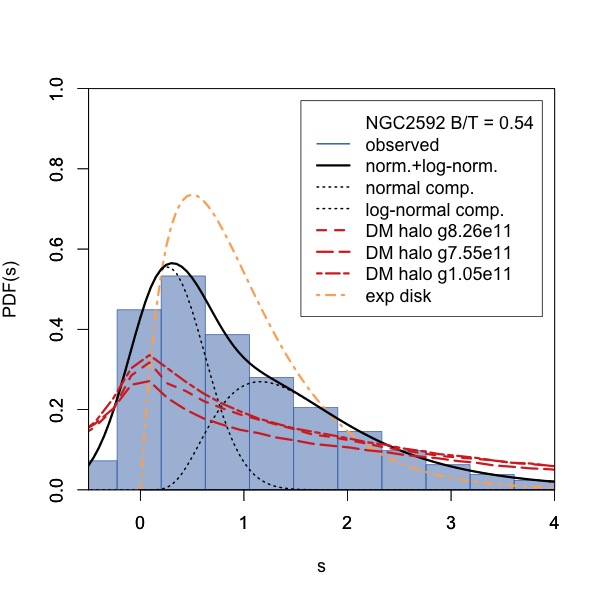}
 \includegraphics[width=0.45\linewidth]{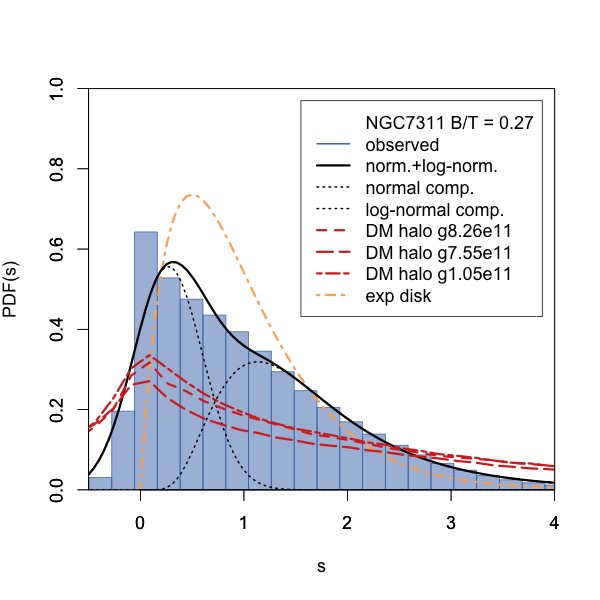}
    \caption{\pdf\/ for three local galaxies in the CALIFA sample, with models and fits overlaid. The blue histogram denotes observed \jj; the black solid curve gives the best-fitting model normal+log-normal function; black dotted curves indicate the normal and log-normal components to that fit; the red {short-dashed, long-dashed and short-long-dashed curves represent the measurement for the dark matter halo in three NIHAO galaxies;} the orange dot-dashed curve is the analytic solution for an exponential disk. The $s<0$ measurements arise from the inclusion of the dispersion term and are physically meaningful.  
    Neither the models nor the fits adequately describe the observations.
 }
   \label{fig:pdfj_o}
\end{center}
\end{figure*}

In Fig.~\ref{fig:pdfj_o} we illustrate a spanning set of local examples highlighted in Fig.~\ref{fig:histograms}. NGC 2906 is a late-type spiral galaxy with low bulge-to-total light ratio $\beta = 0.06$. Its \pdf\/ is broad and symmetric, and peaks near \jj\/ = 1. NGC 2592 is an early-type galaxy with large $\beta = 0.54$; its \pdf\/ is strongly-skewed and peaks near \jj\/ = 0. In between these two extremes, the \pdf\/ for NGC 7311 ($\beta = 0.27$) has characteristics of both disk and bulge. 

\begin{figure*} 
\begin{center}
 \includegraphics[width=0.45\linewidth]{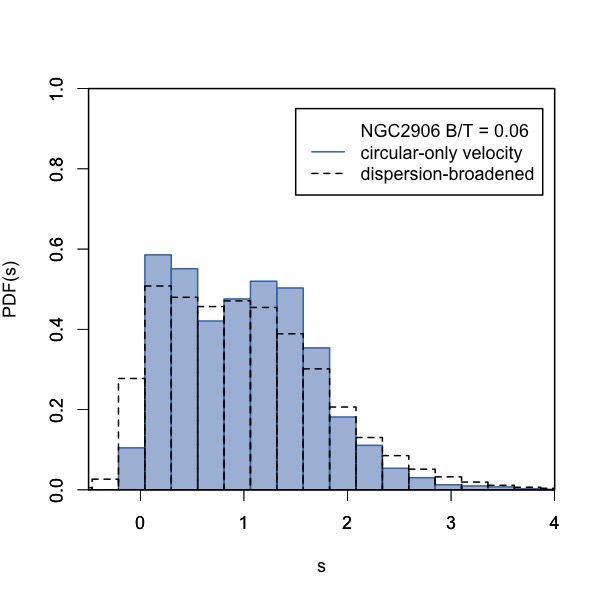}
 \includegraphics[width=0.45\linewidth]{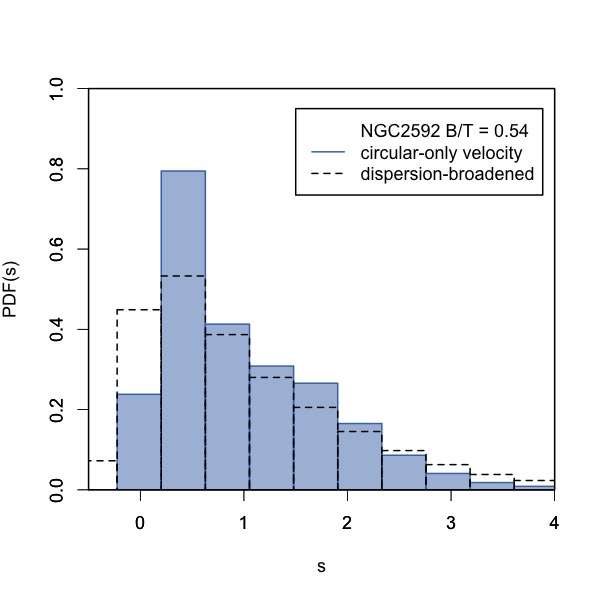}
 \includegraphics[width=0.45\linewidth]{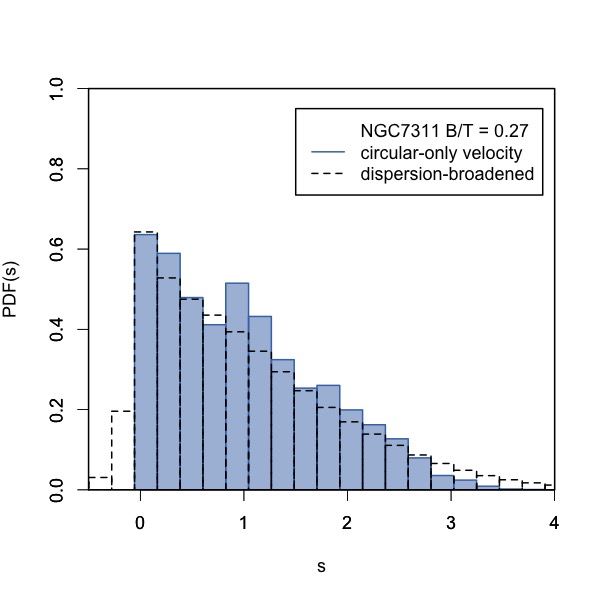}
 \caption{Example \pdf\/ for local galaxies in the CALIFA sample (blue, filled histogram), using circular velocity (not including the dispersion term). The dispersion-broadened \pdf\/ from earlier figures is shown as a black, dashed, open histogram.}
   \label{fig:pdfj_c}
\end{center}
\end{figure*}

In Figure~\ref{fig:pdfj_c} we show \pdf\/ derived using circular-only velocity, that is, calculated without the dispersion term. The circular \pdf\/ for late-type NGC 2906 is symmetric, centred near \jj\/ = 1. For early-type NGC 2592 the circular \pdf\/ peaks nearer \jj\/ = 0.5 and is positively skewed, while NGC 7311 has skewness intermediate between them. In all cases the effect of omitting the dispersion term is to recover more structure in the \pdf\/; in other words, including the dispersion term (as in the dashed histograms and Figure~\ref{fig:pdfj_o}) has the effect of smoothing the \pdf\/. 

\begin{figure} 
\begin{center}
 \includegraphics[width=\linewidth]{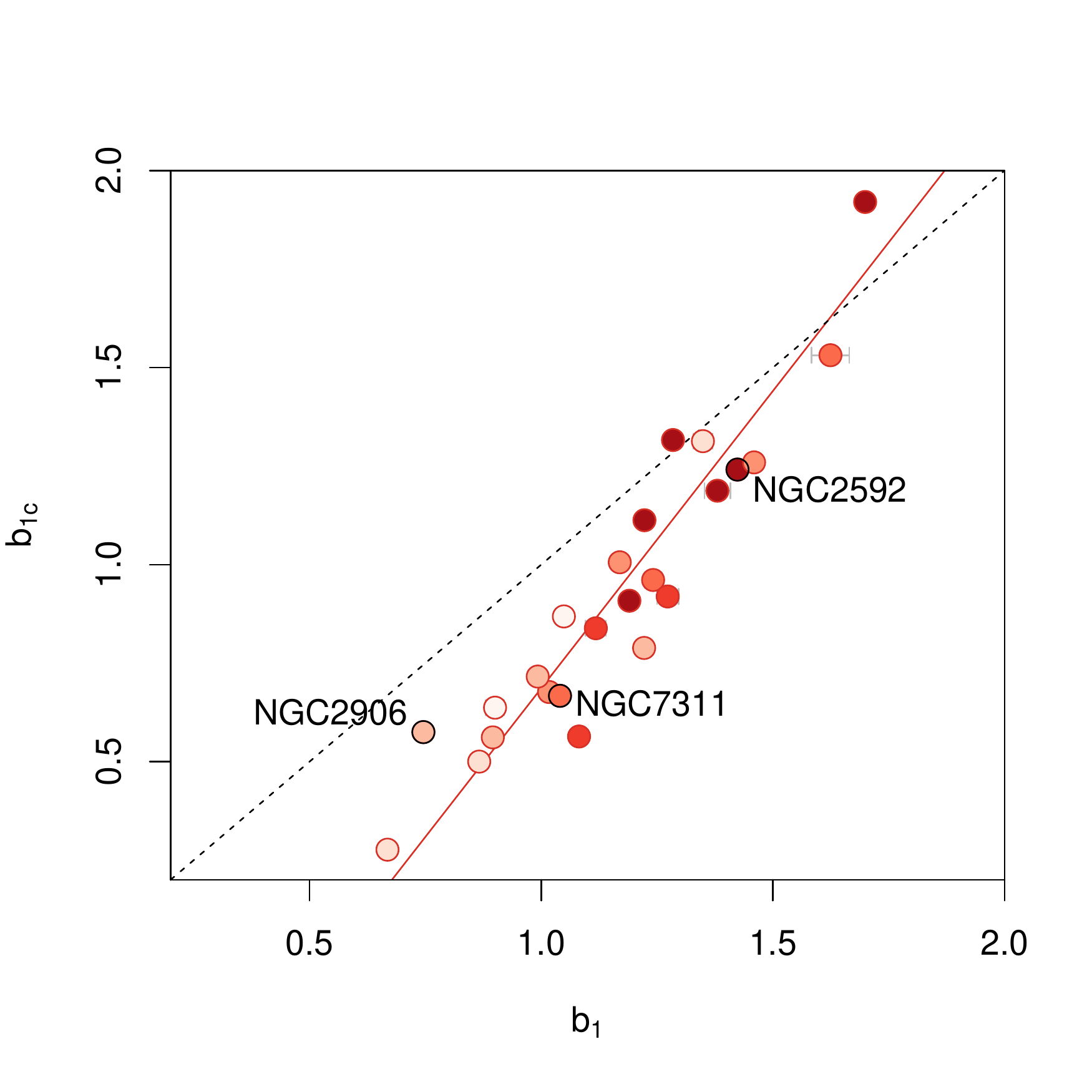} 
 \caption{Correlation between skewness of the bulge \pdf\/ using the method described in Section~\ref{sec:methods} which accounts for velocity dispersion ($b_{1}$) vs. circular-only velocities ($b_{1c}$). { Points are coloured by Hubble type, with redder colours indicating earlier-type galaxies.} The three CALIFA example galaxies presented in Figure~\ref{fig:pdfj_o} are highlighted with black circles and labels. The black, dashed line gives the 1:1 relation. The two methods are tightly correlated, with circular-only \pdf\/ being { on average} less skewed than the dispersion-broadened \pdf\/, { particularly for late-type galaxies which have increasing $\sigma(<{\bf r})$ profiles}. }
   \label{fig:skew_skewc}
\end{center}
\end{figure}

Figure~\ref{fig:skew_skewc} demonstrates the tight correlation between circular \pdf\/ skewness $b_{1c}$ and dispersion-broadened \pdf\/ skewness  $b_{1}$, with correlation coefficient $R=0.94$ and scatter $\sigma = 0.07 \pm 0.01$. Circular \pdf\/ may thus provide a useful comparison in the cases where dispersion measurements are unavailable. {The offset between $b_{1}$ and $b_{1c}$ is a function of Hubble type, with early-type galaxies (darker shades of red) tending to have smaller offsets ($b_{1}~-~b_{1c}$) and late-type galaxies (lighter shades) having larger offsets, with $b_{1} > b_{1c}$. For early-type galaxies, the velocity dispersion profile $\sigma(<{\bf r})$ tends to decrease with radius \citep{Falcon-Barroso+2017} and thus also with $j_*$, so that the effect of including dispersion to calculate $b_1$ has the greatest magnitude at low $s$. Since the bulk of the probability mass is at low $s$, the skewness $b_1$ is not much larger than $b_{1c}$ for the earliest-type galaxies.
Conversely, for late-type galaxies $\sigma(<{\bf r})$ tends to increase with radius and $j_*$, so that the effect of including dispersion has the greatest magnitude at high $s$. The result is that $b_1$ is appreciably larger than $b_{1c}$ for late types. However, we point out that conversion between $b_{1c}$ and $b_1$ can be performed using the coefficients given in Table~\ref{tab:fit}, without needing to measure Hubble type or $\sigma(<{\bf r})$.}

\subsection{Quantifying \pdf\/}
In this subsection we demonstrate and explain that simple models and various functional forms do not well describe \pdf\/.

Motivated by the fact that the naive model where $j_b(disk) \sim j_{halo}$ works surprisingly quite well, we also explore a naive model where \pdf\/ $\sim$ \pdfh\/ 
\footnote{This comparison is similar to the \citet{vandenBosch+2001} analysis, except that was concerned with the norm of $j$ rather than the $z-$component we show in this work:
\begin{equation}
\label{eq:vdb}
p(s)=\frac{\zeta \mu(\mu-1)}{(\zeta s+\mu-1)^{2}}
\end{equation}
where $\mu$ is a parameter describing the mass distribution of $j/j_{\text {max}}$ and 
\begin{equation}
\zeta=\frac{j_{\text {mean}}}{j_{\max }}=1-\mu\left[1-(\mu-1) \ln \left(\frac{\mu}{\mu-1}\right)\right].
\end{equation}
They assumed $\mu = 1.25$, the average found by \citet{Bullock+2001}, though we note that Eqn~\ref{eq:vdb} is peaked at $s=0$ for any value of $\mu$.
They investigated a sample of dwarf galaxies and similarly found a lack of low-\jj\/ material, as well as of high-\jj\/ material. Indeed, \citet{SS05} found that Equation~\ref{eq:vdb} does not well describe an NFW halo with an embedded exponential disk, regardless of choice of parameter $\mu$, or model concentration, virial radius, scale length or mass fraction.}. Our {DM controls come from the NIHAO simulations \citep{Wang+2015}, where halo \pdfh\/ is measured for the DM inside the virial radius at $z=0$, and the $z$-direction is defined as the direction of the mean angular momentum vector of all matter inside the virial radius. The individual particle values were used in order to align with our observational method of accounting for the dispersion. 
Three NIHAO galaxies were selected to approximately span the mass range of our CALIFA sample and to qualitatively demonstrate that the NIHAO \pdfh\/ tend to be rather self-similar, but different to the observed \pdf\/ for CALIFA galaxies. NIHAO galaxy g8.26e11 is a Milky-Way-like galaxy, with virial mass $1.02 \times 10^{12}M_\odot$ and stellar mass $4.68 \times 10^{10}M_\odot$ 
; Galaxy g7.55e11 has virial mass $8.93 \times 10^{11}M_\odot$ and stellar mass $3.11 \times 10^{10}M_\odot$; 
Galaxy g1.05e11 is a dwarf galaxy, with virial mass $1.18 \times 10^{11}M_\odot$ and stellar mass $5.67 \times 10^{8}M_\odot$ 
at $z=0$ \citep{Wang+2015}. 
See \citet{Wang+2018} for further analysis of g8.26e11. The \pdfh\/ for these galaxies is shown by the red, short-dashed, long-dashed and short-long-dashed lines in Fig.~\ref{fig:pdfj_o}. It is clear that the halo curves do} not at all describe late-type NGC 2906, intermediate NGC 7311, or even early type NGC 2592. In all cases there is less baryonic material compared with the DM halo at low \jj\/ ($\lesssim0$), more at average \jj\/ ($\sim 1$), and less at high \jj\/ ($\gtrsim3$).

The analytic solution for an exponential disk with constant rotation velocity is
\begin{equation}
\label{eq:exp}
p(s)=4s {\rm e}^{-2s},
\end{equation}
plotted in Fig.~\ref{fig:pdfj_o} as orange dot-dashed lines. As expected, this model fails to describe the galaxies in our sample, even the almost-pure disks such as NGC 2906 which has $\beta=0.06$, since the rotation velocities are not constant but decrease towards the galaxy centres.

In attempt to better parametrize the \pdf\/, \citet{SS05} fitted a Gamma distribution function\footnote{Note that the \citet{SS05} work was also concerned with the norm of $j$ rather than the $z$-component.}
\begin{equation}
P(j)=\frac{1}{j_{d}^{\alpha_\Gamma} \Gamma(\alpha_\Gamma)}j^{\alpha_\Gamma-1} e^{-j / j_{d}},
\end{equation}
where
\begin{equation}
j_{d}=\frac{j_{\text {mean}}}{\alpha_\Gamma},
\end{equation}
which is described by a single shape parameter $\alpha_\Gamma$. The function goes to zero at \jj\/ = 0 when $\alpha_\Gamma > 1$ and to $\infty$ when $\alpha_\Gamma <1$. A value of $\alpha_\Gamma = 2$ describes an exponential disk embedded in an isothermal halo with constant rotation velocity.
\citet{SS05} showed that gas has higher $\alpha_\Gamma$ than the halo because the bulk of its mass rotates more quickly in the inner parts { for the `particle method' of analysis, though no significant difference was found between the gas and halo for their other methods}. 
However, since the fit is critically unstable at $\alpha_\Gamma=1$ and is strongly biased by the few bins around \jj$\sim 0$ and the choice of bin width, the fit often fails to converge for galaxies in our sample, and is consequently not plotted in Fig.~\ref{fig:pdfj_o}.

\begin{figure} 
\begin{center}
 \includegraphics[width=\linewidth]{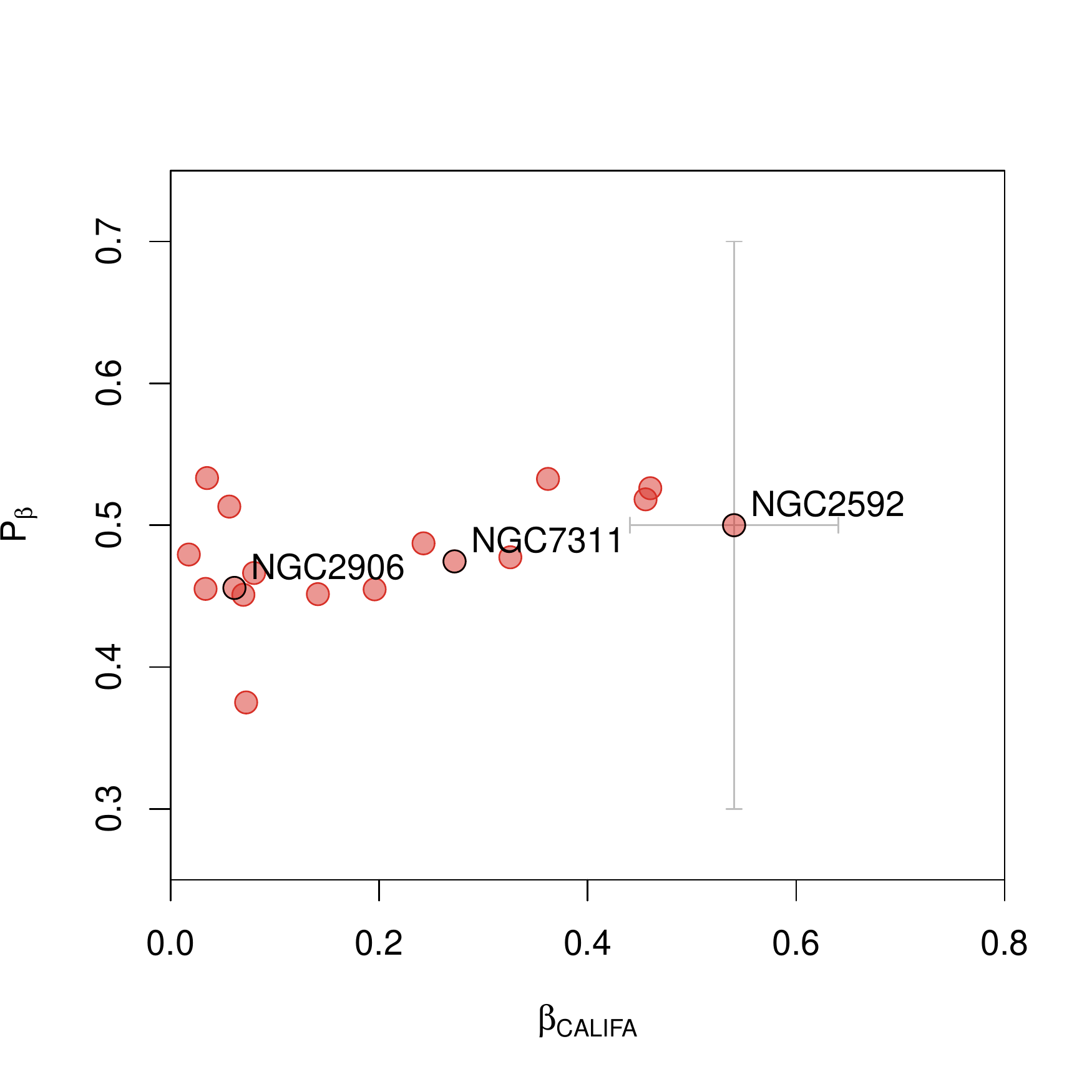}
    \caption{Comparison between {fractional} area under the Gaussian curve illustrated in Fig~\ref{fig:pdfj_o}, against bulge-to-total ratio from CALIFA. Three example galaxies presented in Fig.~\ref{fig:pdfj_o} are highlighted with black circles. A typical error bar is shown. The fitted $P_\beta$ does not correlate with CALIFA $\beta$.
 }
   \label{fig:bt_nln}
\end{center}
\end{figure}

Instead we attempt to fit the \pdf\/ with a simple, two-component model intended to describe the dispersion-dominated material (that is, bulge, thick disk, clumps) and the rapidly-rotating thin disk. The random motion of the dispersion-dominated material can be approximated with a normal distribution that peaks at \jj\/~=~0, allowing a component of negative velocity material as demonstrated in \citet{vandenBosch+2002}\footnote{If one considers that the dispersion-dominated material may be skewed towards positive \jj\/, the normal distribution could be replaced with a second log-normal distribution, but we found that such fits were not well constrained.}.
Ideally the thin disk would be represented by Equation~\ref{eq:exp}, which assumes an exponential disk surface brightness with scale length $r_d$ and a constant rotation velocity $v_{\rm max}$; alternatively a more typical rotation curve could be assumed, e.g. ${v(r)} = v_{\rm max}(1-{\rm exp}(-{r}/r_{\rm flat}))$. We found that the combined normal + Eqn.~\ref{eq:exp} fits frequently failed to converge in our sample as they do not well describe even the high-\jj\/ part of the distributions. The typical rotation curve model is less simple, since it varies with both $r_d$ and $r_{\rm flat}$; it therefore introduces additional variation with disk concentration $r_d/r_{\rm flat}$ which is not constant with stellar mass \citep{OG14}. {Consequently, the corresponding analytic solution $p(s)$ is not a simple function of $s$ alone but also of covariant parameters $v_{\rm max}$ and $r_{\rm flat}$. Such fits also failed to converge for our sample.} 
We thus settle on a normal + log-normal distribution. We perform these fits using a non-linear least squares approach on the unbinned data, using the Levenberg-Marquardt algorithm \citep{More1978}.
The normal + log-normal distribution fits are overlaid in black on the examples in Figure~\ref{fig:pdfj_o}. These distributions tend to fit the \pdf\/ reasonably well at $s>1$ but less so at $s<1$. For example, for NGC 7311 the peak near $s\sim0$ is not well captured by the normal curve. Further, one could define the area under the normal, `bulge' component curve to be $A_\beta$, and the area under the normal + log-normal curve as $A$. One would then expect the ratio $A_\beta/A = P_{\beta} $, to be related to bulge-to-total ratio $\beta$. However, Figure~\ref{fig:bt_nln} illustrates that $P_{\beta} $ does not correlate with $ \beta_{\rm CALIFA}$, so we conclude that the normal + log-normal fit is not a useful method for quantifying the \pdf\/ and its relationship with galaxy photometric morphology.

In sum, neither the analytical CDM fits nor the normal+lognormal model fits describe the data well for the purposes of this work. Instead it seems that the best method for quantifying \pdf\/ is with the statistics of the underlying distribution as defined in Section~\ref{sec:methods}, particularly skewness $b_1$. These statistics are robust against choice of bin size and are model-independent, so are our choice for quantifying \pdf\/ in the remainder of the paper.

\section{The \pdf\/ - morphology relation} 
\label{sec:analysis}

In this section we analyse the relation between the shape of \pdf\/, quantified by skewness $b_1$, and galaxy morphology, quantified by bulge-to-total ratio $\beta$ and Hubble type $T$. We also investigate \pdf\/ calculated without dispersion, and bulge-disk decomposition.

\begin{figure*} 
\begin{center}
 \includegraphics[width=0.49\linewidth]{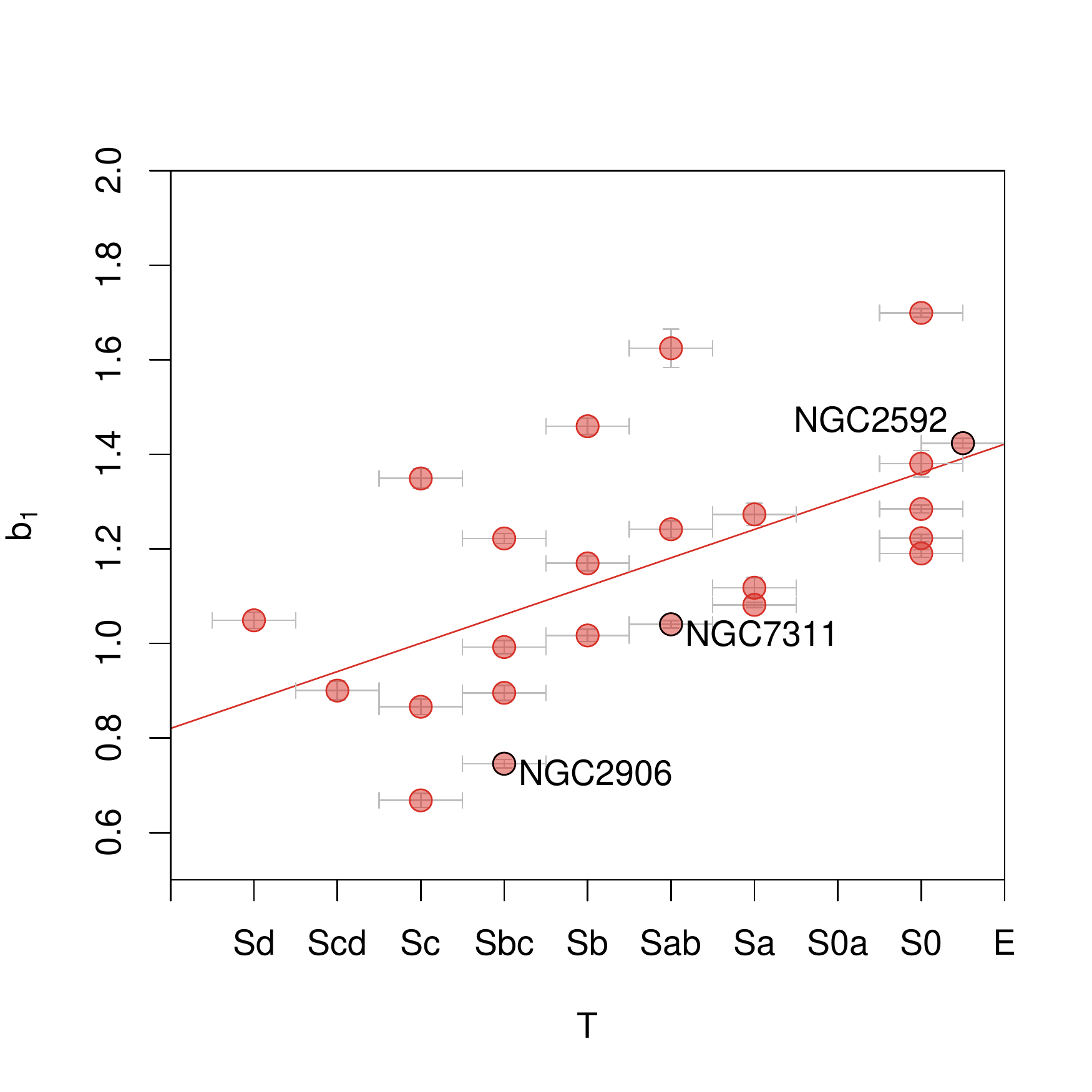} 
  \includegraphics[width=0.49\linewidth]{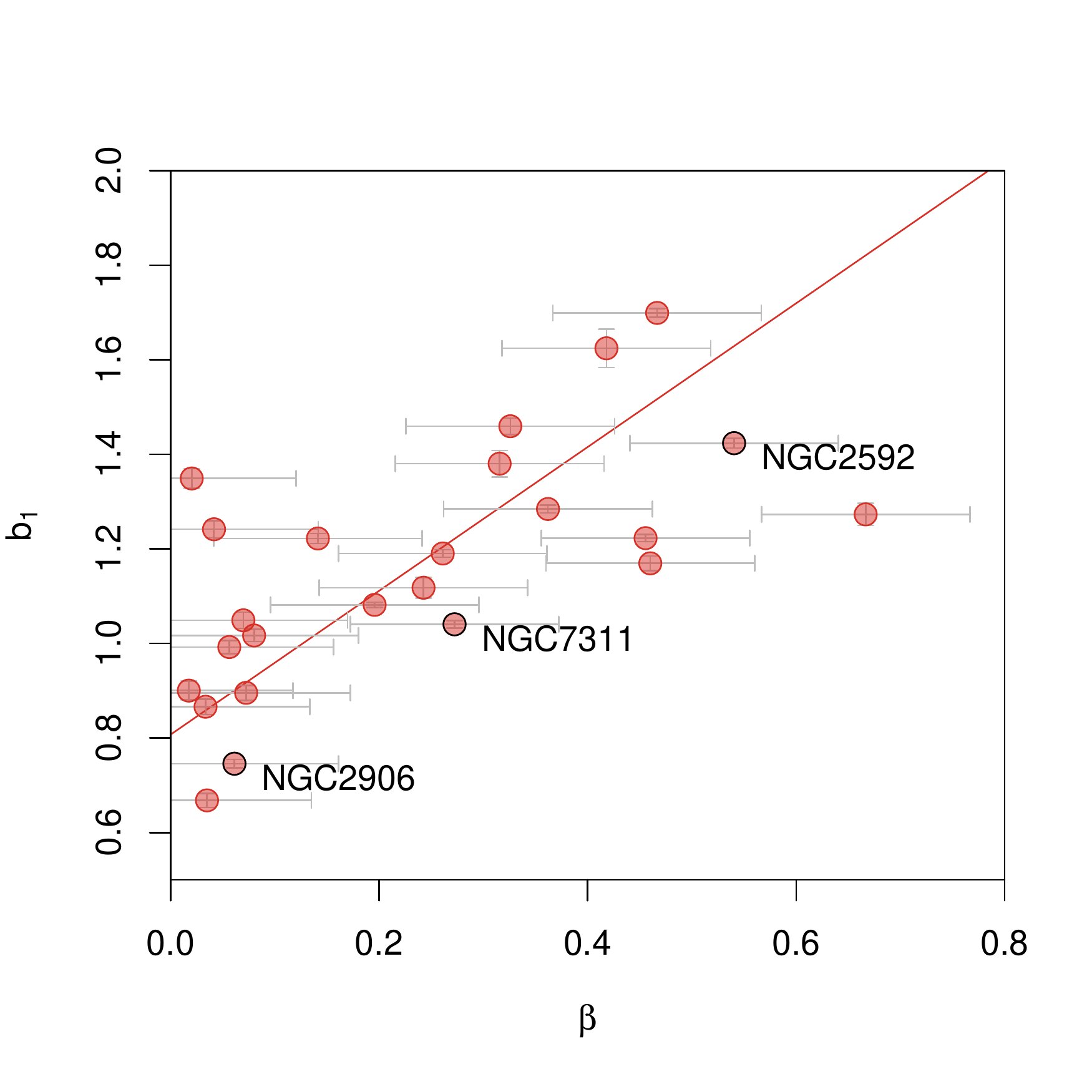} 
 \caption{Correlation between morphology and shape of \pdf\/ quantified by skewness $b_1$. The three CALIFA example galaxies presented in Figure~\ref{fig:pdfj_o} are highlighted with black circles and labels. Left: \pdf\/ skewness $b_1$ is positively correlated with Hubble type $T$.
 Right: $b_1$ is positively correlated with bulge-to-total light ratio $\beta$. Earlier-type galaxies with bigger bulges have \pdf\/ that are more strongly-skewed. }
   \label{fig:t-kurt}
\end{center}
\end{figure*}

Recall that Figure~\ref{fig:histograms} illustrates a correlation between \pdf\/ shape, $\beta$ and $T$, where late-type galaxies with low $\beta$ have symmetric \pdf\/ with skewness $b_1 \sim 0.8$, while early-type galaxies with high $\beta$ have strongly positively-skewed \pdf\/ with $b_1 > 1.2$. In Figure~\ref{fig:t-kurt} we show the correlation between $T$ and $b_1$, and between $\beta$ and $b_1$; the corresponding fit parameters and correlation coefficients \citep[derived using \texttt{hyper.fit} from][]{Robotham+2015} are given in Table~\ref{tab:fit}. The Hubble type units are such that the difference between two tick-marks $\Delta T = Sd - Scd = 1$. We find that earlier-type galaxies that have bigger bulges tend to have \pdf\/ that are more strongly-skewed. There is a moderate correlation between $T$ and $b_1$, with Pearson correlation coefficient $R=0.58$ and significant scatter $\sigma = 0.20 \pm 0.03$. The correlation between $\beta$ and $b_1$ is stronger, with $R=0.66$ and $\sigma = 0.09 \pm 0.02$. The fact that the $\beta-b_1$ correlation is stronger than the $\beta-T$ relation may reflect that Hubble type classification is more subjective than bulge-to-total light ratio. {We suspect that t}he reason behind the \pdf\/ - morphology relation is that the processes that cause the build-up, ejection and rearranging of angular momentum are those that contribute to the build-up of the bulge, effecting morphological change towards earlier types. The skewness of \pdf\/ can therefore be used as a quantitative morphological tracer that is based on fundamental properties.

We do not find a strong correlation between morphology and kurtosis $b_2$, so have not shown the corresponding fits here.

\begin{figure*} 
\begin{center}
 \includegraphics[width=0.49\linewidth]{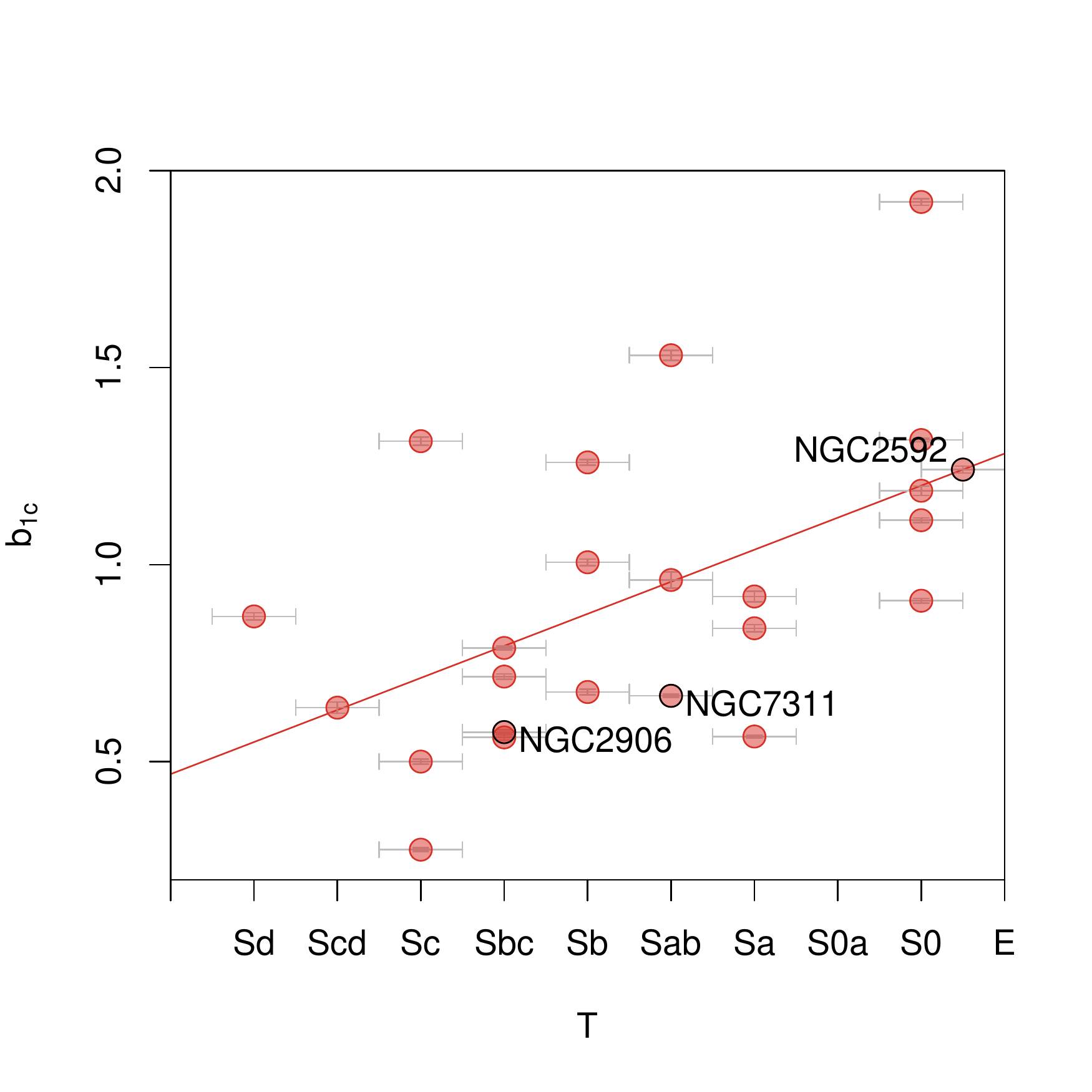} 
  \includegraphics[width=0.49\linewidth]{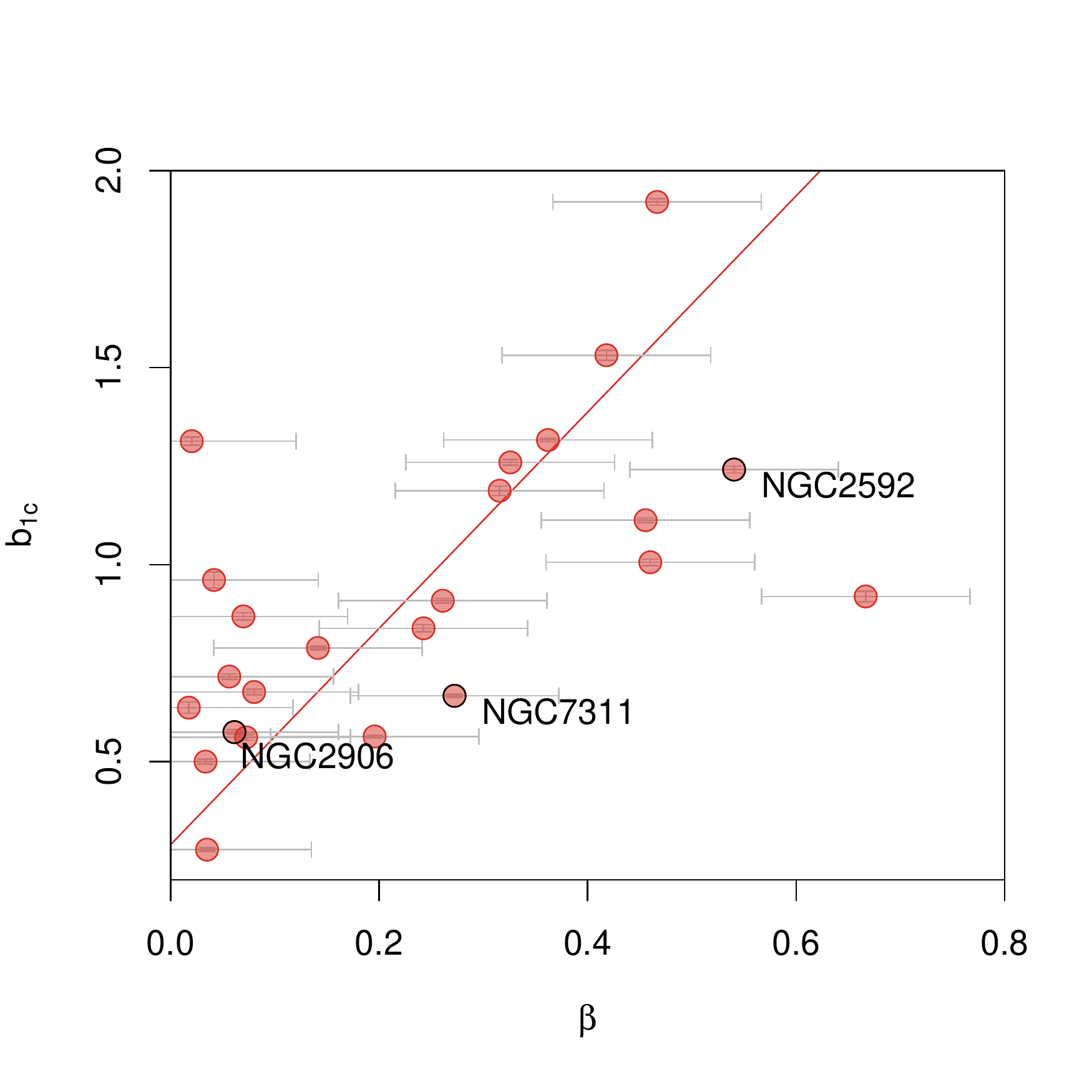} 
 \caption{Correlation between morphology and shape of circular \pdf\/ quantified by skewness $b_{1c}$. The three CALIFA example galaxies presented in Figure~\ref{fig:pdfj_o} are highlighted with black circles and labels. Left: \pdf\/ skewness $b_{1c}$ is positively correlated with Hubble type $T$.
 Right: $b_{1c}$ is positively correlated with bulge-to-total light ratio $\beta$. Earlier type galaxies with bigger bulges have circular \pdf\/ that are more strongly-skewed.}
   \label{fig:t_skewc}
\end{center}
\end{figure*}

Figure~\ref{fig:t_skewc} illustrates that the skewness of circular-velocity \pdf\/ $b_{1c}$ correlates with Hubble type $T$ and bulge-to-total ratio $\beta$. The correlations are similar in strength to the corresponding correlations with $b_1$. At first glance, the fact that $b_{1c}$ correlates with morphology may seem unsurprising given the tight correspondence between $b_{1c}$ and $b_{1}$, demonstrated earlier in Figure~\ref{fig:skew_skewc} and Table~\ref{tab:fit}. However, it is non-trivial that omitting the fundamental morphological history that is encoded in the stellar velocity dispersion does not strongly affect the correlation with morphology. Likewise it is noteworthy that if dispersion measurements are unavailable (e.g. at low spectral resolution, of gas kinematics only, and/or at high redshift), then one can still use circular velocities to obtain a meaningful \pdf\/-based measure of morphology.

\begin{table} 
\centering
\caption{Fit parameters for skewness of \pdf\/ against Hubble type, bulge-to-total ratio,  and bulge S{\'ersic} index.} 
\label{tab:fit}
\begin{tabular}{lrrrrrrr}
\hline
$x$ & $a$ & $\Delta a$ & $b$ & $\Delta b$ & $\sigma$ & $\Delta \sigma$ & $R$\\
\hline
$y=b_1$\\
$T$ (Fig.~\ref{fig:t-kurt}a) & 0.82 & 0.10 & 0.06 & 0.02 & 0.20 & 0.03 & 0.58\\
$\beta$ (Fig.~\ref{fig:t-kurt}b) & 0.81 & 0.09 & 1.52 & 0.32 & 0.09 & 0.02 & 0.66\\
\hline
$y=b_{1c}$\\
$b_1$ (Fig.~\ref{fig:skew_skewc}) & -0.82 & 0.13 & 1.51 & 0.11 & 0.07 & 0.01 & 0.94\\
$T$ (Fig.~\ref{fig:t_skewc}a) & 0.47 & 0.16 & 0.08 & 0.03 & 0.31 & 0.05 & 0.53\\
$\beta$ (Fig.~\ref{fig:t_skewc}b) & 0.29 & 0.18 & 2.74 & 0.69 & 0.11 & 0.03 & 0.60\\
\hline
$y=b_{1rb}$\\
$T$ (Fig.~\ref{fig:t_skewrb}a) & 0.43 & 0.10 & 0.03 & 0.02 & 0.17 & 0.03 & 0.41\\
$\beta$ (Fig.~\ref{fig:t_skewrb}b) & 0.35 & 0.05 & 1.03 & 0.17 & 0.05 & 0.02 & 0.77\\
$n_b$ (Fig.~\ref{fig:bulgen_kurtrb}) & 0.34 & 0.06 & 0.13 & 0.03 & 0.12 & 0.02 & 0.67\\
\hline
\end{tabular}\\
$y = a + bx$. $T$ denotes Hubble Type, $\beta$ is the bulge-to-total light ratio, and $n_b$ is the bulge S{\'e}rsic index.   The $\Delta$ prefix denotes measurement uncertainty on each parameter, while  $\sigma$ is intrinsic scatter along the $y$-axis and $R$ is the Pearson correlation coefficient.
\end{table}

\subsection{Bulge-disk decomposition}

\begin{figure*} 
\begin{center}
\vspace{-12pt}
 \includegraphics[width=0.45\linewidth]{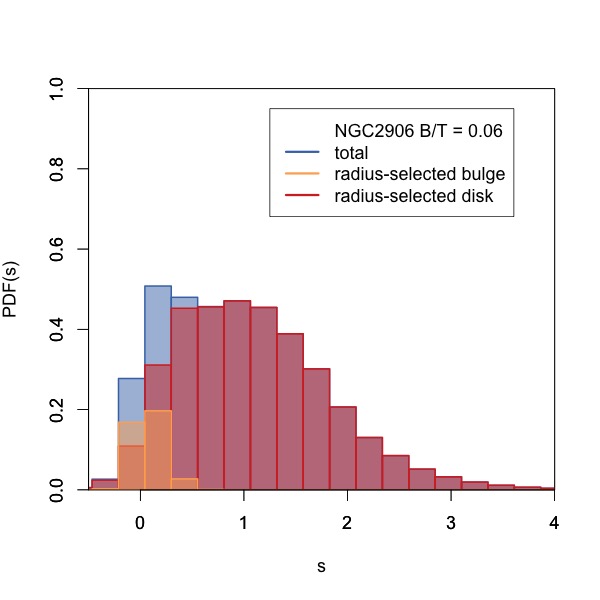}
  \includegraphics[width=0.45\linewidth]{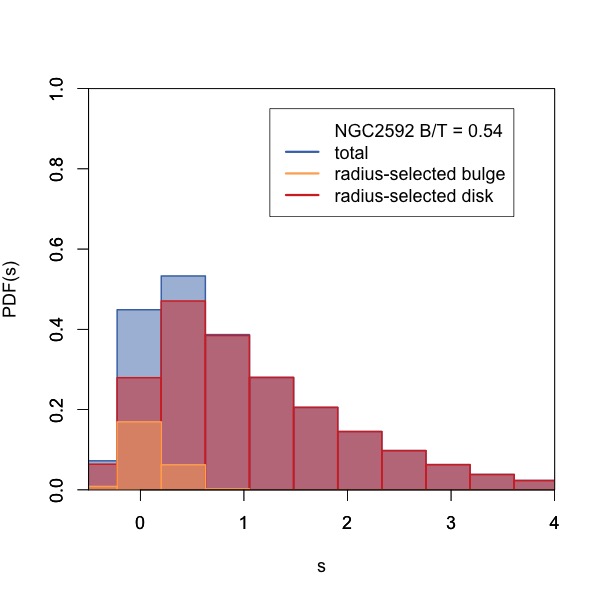}
 \includegraphics[width=0.45\linewidth]{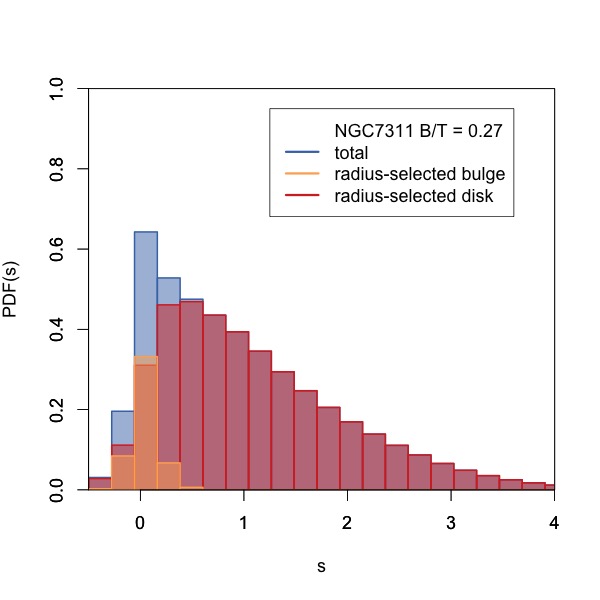}
  \caption{Example \pdf\/ for local galaxies in the CALIFA sample (blue), with bulge (orange) and disk (red), using radius-based selection. See text for details of calculation.}
   \label{fig:pdfj_r}
\end{center}
\end{figure*}

In Figure~\ref{fig:pdfj_r} we show \pdf\/ for NGC~2906, NGC~2592 and NGC~7311 decomposed into `bulge' and `disk' components. 
We use CALIFA $\beta$ to determine the fraction of light to assign to each component under a radius-based selection. We sum up the spaxels representing the cylinder of material nearest to the galaxy centre, increasing the radius of the cylinder until its mass (surface density) as a fraction of the total equals $\beta$. These spaxels are then assigned to the radius-selected bulge.  Their \pdf\/ is coloured orange in this figure and its skewness is denoted $b_{1rb}$. The remaining spaxels are assigned to the radius-selected disk; their \pdf\/ is coloured red. The component distributions are normalised to the mean $j_*$ of the galaxy rather than the mean $j_*$ of each component. {The radii of the bulges selected in this manner are in all cases larger than the PSF half-width at half-maximum of 1.85\arcsec and larger than the central spaxel size of 1\arcsec \citep{Husemann+2013}, so  we do not expect this analysis to be greatly affected by beam-smearing.}

\begin{figure*} 
\begin{center}
 \includegraphics[width=0.49\linewidth]{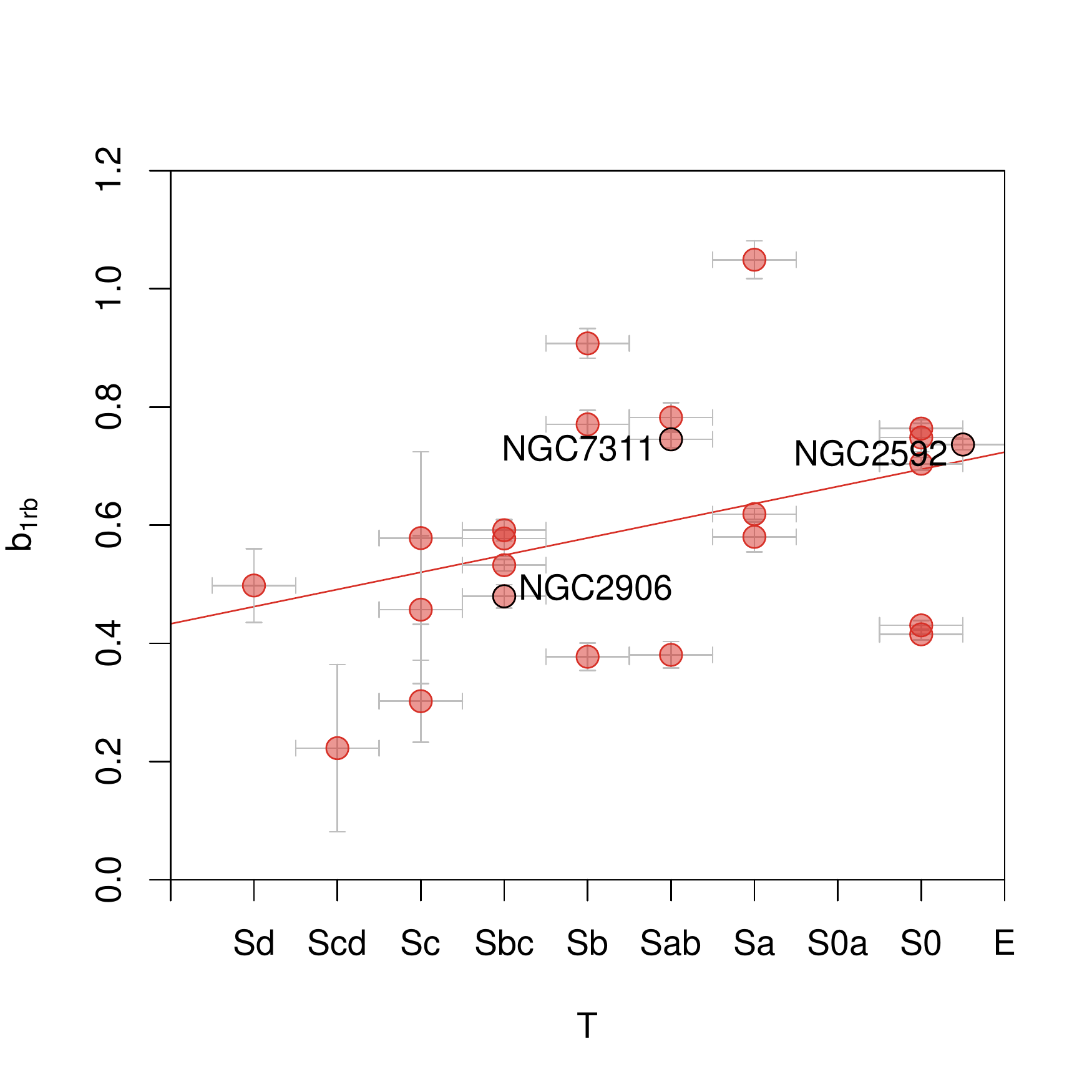} 
  \includegraphics[width=0.49\linewidth]{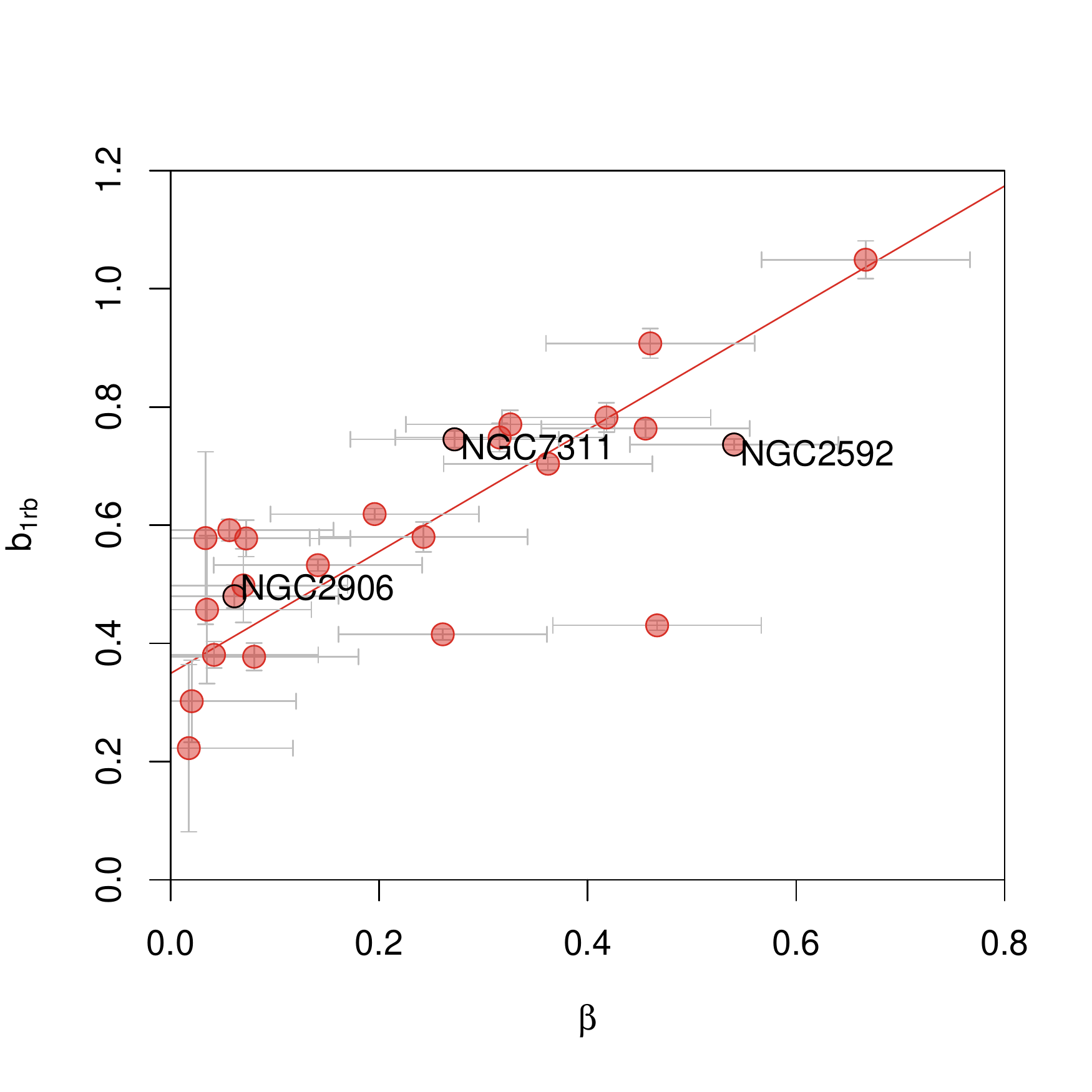} 
 \caption{Correlation between morphology and shape of bulge \pdf\/ quantified by skewness $b_{1rb}$. The three CALIFA example galaxies presented in Figure~\ref{fig:pdfj_o} are highlighted with black circles and labels. Left: bulge \pdf\/ skewness $b_{1rb}$ is correlated with Hubble type $T$.
 Right: $b_{1rb}$ is correlated with bulge-to-total light ratio $\beta$. The angular momentum distributions of bigger bulges in earlier-type galaxies are more strongly skewed.}
   \label{fig:t_skewrb}
\end{center}
\end{figure*}

The radius-selected bulge \pdf\/ tend to be moderately positively-skewed, with $0.2 < b_{1rb} < 1.1$.
In Figure~\ref{fig:t_skewrb} we show that $b_{1rb}$ is correlated with Hubble type and with $\beta$, where the bulges in early-type galaxies tend to have \pdf\/ that are more strongly skewed.

\begin{figure} 
\begin{center}
 \includegraphics[width=\linewidth]{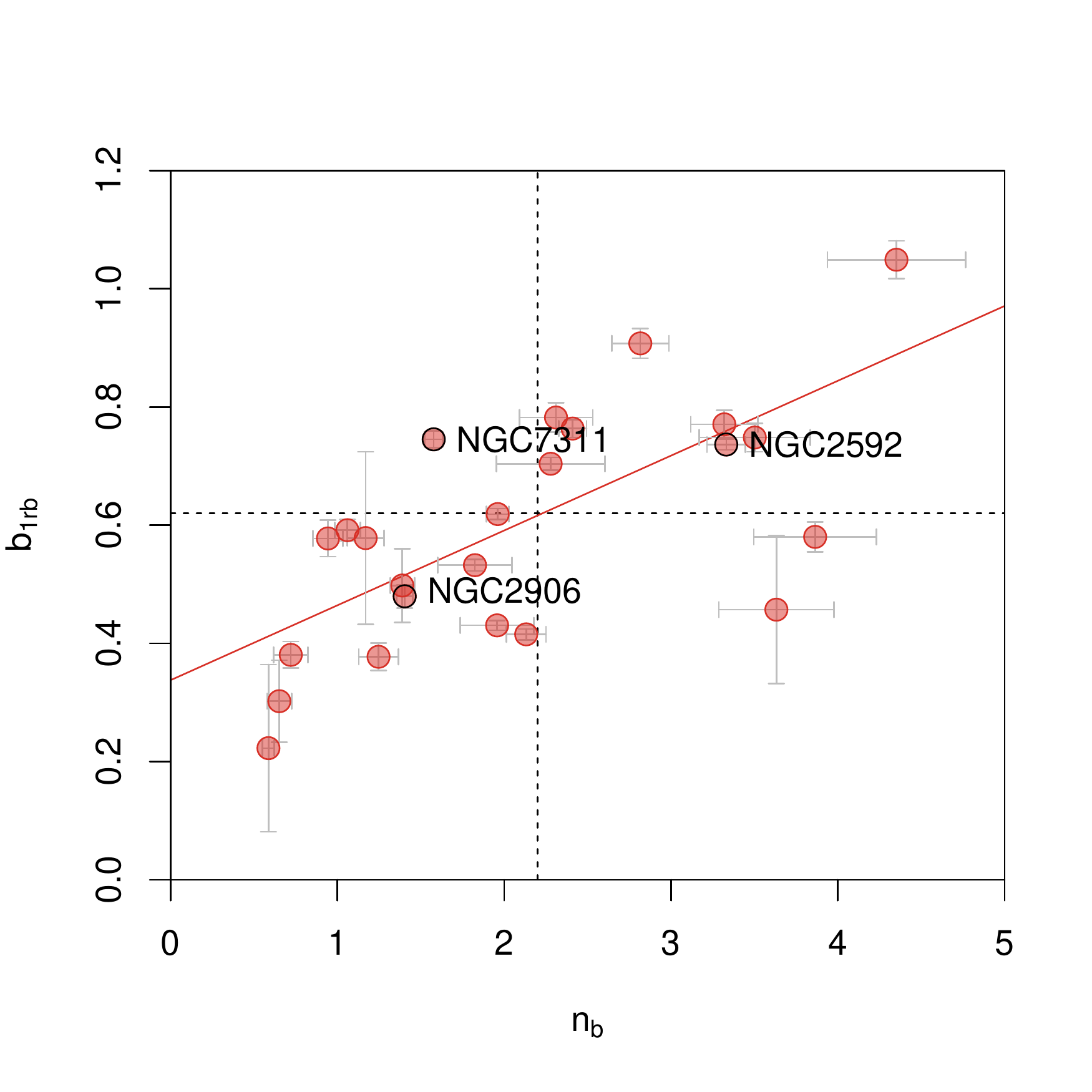} 
 \caption{Correlation between CALIFA bulge S{\'e}rsic index $n_b$ and skewness of radius-selected bulge \pdf\/ $b_{1rb}$. The vertical, dashed line shows a pseudobulge cut, where $n_b<2.2$ corresponds to galaxies that host pseudobulges, and $n_b\geq2.2$ includes galaxies that may host pseudobulges or classical bulges. The horizontal, dashed line shows a similar cut in $b_{1rb}$, where almost all of the galaxies that are known to host pseudobulges are below $(b_{1rb})>0.62$. Bulges with higher S{\'e}rsic index have more strongly skewed \pdf\/.}
  \label{fig:bulgen_kurtrb}
\end{center}
\end{figure}

Figure~\ref{fig:bulgen_kurtrb} investigates the relationship between bulge S{\'e}rsic index \citep{Sersic1963} and the skewness of the radius-selected bulge \pdf\/ $b_{1rb}$. There is a clear correlation, such that bulges with higher S{\'e}rsic index have more strongly skewed \pdf\/. The vertical, dashed line corresponds to the \citet{fisher2008,fisher2016} 
diagnostic for bulge type, where galaxies that contain pseudobulges have bulge S{\'e}rsic index $n_b<2.2$, and galaxies with $n_b\geq2.2$ may host pseudobulges or classical bulges. 
We note that almost all of the galaxies known to host pseudobulges {lie below $b_{1rb} \lesssim 0.62$, which is the skewness at which the pseudobulge diagnostic intersects the best-fitting relation between $b_{1rb}$ and $n_b$}.
We suggest that $b_{1rb}$ may thus also be related to bulge type, with bulges $b_{1rb}\lesssim0.62$ corresponding to pseudobulges and more strongly skewed bulges having $b_{1rb}\gtrsim0.62$ corresponding to classical bulges. { This is a similar correlation to that for total galaxy skewness $b_1$; in both 
cases, rotation-dominated systems (pseudobulges and late-type galaxies) have normally-distributed $s$, and dispersion-dominated systems (classical bulges and early-type galaxies) are strongly-skewed. One may naively expect the opposite, where it is the dispersion-dominated systems that have normally-distributed $s$; we suspect that their enhanced skewness may be a result of angular momentum being rearranged as the bulge is built up, where material of average $s$ loses angular momentum, leaving behind a tail towards high-$s$. 
The positive correlation between $b_{1rb}$ and $n_b$ appears to be} related to the finding by \citet{fabricius2012} that there is a correlation between bulge S{\'e}rsic index and velocity dispersion profile, in that a large $n_b$ indicates a central component with higher velocity dispersion than the disk.


Clearly this is simply a first demonstration of such a \pdf\/-based decomposition, using photometric $\beta$ as a prior. A more sophisticated decomposition would ideally employ an iterative approach to arrive at a photokinematic $\beta$. This would allow to test to what extent $b_{1rb}$ could be used to classify bulge type. It would also be interesting in future work to explore the effect of spatial resolution. 



\section{Discussion} 
\label{sec:discussion}

We have shown that the spatially-resolved stellar specific angular momentum distribution, \pdf\/, of a morphologically-broad sample does not match {a selection of toy} analytic solutions and simple functions. Instead, we characterise the shape of the \pdf\/ with its skewness $b_1$, which correlates with photometric morphology. In this section we discuss the implications of our findings in the context of galaxy evolution.

In Figure~\ref{fig:pdfj_o} we demonstrated that the observed stellar \pdf\/ differs significantly from the corresponding \pdf\/ of constantly rotating exponential disks, as well as the \pdfh\/ expected for DM haloes\footnote{Neither model is expected to fit precisely, since the rotation density profiles of baryonic disks differ from those of DM, and rotation curves are not flat at the centres of galaxies.}. A normal plus log-normal distribution also fails to fit the range of \pdf\/ shapes. 
Thus even though the baryonic $j_b$ is shown to be empirically similar to halo $j_h$  \citep{Barnes+1987,Catelan+1996a,Catelan+1996b,vandenBosch+2001,Posti2018MN}, which appears to be something of a coincidence, we find that the stellar \pdf\/ is completely different to the halo \pdfh\/. 
The difference between the observed and predicted distributions shows that galaxy assembly is dominated by angular momentum transport processes experienced by the baryons after decoupling from the DM halo, and the fine tuning between $j_b$ and $j_h$ probably \emph{is} a coincidence. 

Compared with the DM halo models there is a dearth of low-\jj\/ (\jj\/ $\lesssim 0$) baryonic material, an excess of average-\jj\/ $\sim 1$ material, and a smaller lack of high-\jj\/ (\jj\/ $\gtrsim 3$) material across the full range of morphological types. Processes that selectively decrease angular momentum within a galaxy include outflows and tidal stripping \citep{vandenBosch+2001}. Low-\jj\/ material may be ejected by stellar- and AGN-feedback; such outflows preferentially have low angular momentum as predicted by \citet{Brook+2011} and confirmed by \citet{Sharma+2012}.  
The amount of material lost due to outflows is dependent on stellar mass, since the deeper potential of high-mass systems may cause the outflowing material to fall back onto the galaxy in a galactic fountain and be redistributed throughout the disk \citep{Brook+2012}. The large discrepancy between observed and predicted distributions in the low-\jj\/ regime indicates that outflows may be important in the evolution of normal local galaxies such as in the CALIFA survey.  
High-\jj\/ material can be stripped when galaxies interact with local tidal fields, removing rapidly-rotating material at the outskirts \citep{vandenBosch+2001}. The fact that there is only a small difference between the observed and predicted \pdf\/ at the high-\jj\/ regime indicates that tidal stripping is not an important process for these galaxies; this is consistent with their typically isolated environment within the local universe. 

However, neither outflows nor stripping can explain the apparent \emph{excess} of baryonic material above that predicted at moderate \jj\/.  Instead, the height of the distribution suggests that either the baryons experience a redistribution of angular momentum from the initial state when it was coupled with its DM halo, or the baryon fraction in the disk is small and the baryon distribution was initially higher than that of the DM. Angular momentum could be redistributed via viscosity, but that would cause low-\jj\/ material to fall to the centre and high-\jj\/ material to be pushed to the outskirts, increasing the amount of low- and high-\jj\/ material, which is not what is seen here. Alternative mechanisms include redistribution of angular momentum via high-dispersion components such as star-forming clumps, spiral arms and bars, or by merging with nearby galaxies. Simulations show that accreted gas tends to have higher $j$ than the DM halo \citep{Stevens+2017,Stewart+2017,El-Badry+2018}. Modern theory predicts that disc contraction, or biased collapse, is also important in explaining the difference between angular momentum distributions of the DM halo and the baryons \citep{Dekel+2014,Posti2018MN}.

The shape of the \pdf\/ is parametrized by higher-order moments, where $b_1$ quantifies the skewness of the distribution. We showed that this parameter correlates with the photometric morphological tracers Hubble type $T$ and bulge-to-disk light ratio $\beta$. For each of these relations there is non-zero intrinsic scatter around a linear fit when accounting for uncertainties in both axes. Arguably, \pdf\/, which is computed from both kinematic and photometric data, encodes more physical information than photometry alone, so may be a more robust tracer of galaxy type. To test this suggestion would require 1) careful photometric classification of a larger sample of galaxies, e.g. the SAMI Galaxy Survey \citep{Bryant+2015}, with forthcoming bulge-disk decompositions using ProFit \citep{Robotham+2017} by Sarah Casura et al. (in prep), and Stefania Barsanti et al. (in prep), or 2) simulations such as Magneticum \citep{Teklu+2015,Schulze+2018} or EAGLE \citep{Schaye+2015,Lagos+2017,Lagos+2018}, as far as these are well-matched to observations \citep{vandeSande+2019}, which would enable investigation of the strength of correlation with other fundamental properties. 

Interestingly, while $b_1$ traces morphology, {as does global or aperture $j_*$ \citep[e.g.][]{Fall1983,RF12,OG14,Cortese+2016,Sweet+2018}, the same cannot be said for the cousin spin parameter, either global, aperture, or PDF. Here spin parameter $\lambda = J | E |^{1/2} / GM^{5/2}$ \citep{Peebles1969}, with $J$ total angular momentum, $E$ energy, $M$ mass, and $G$ Newton's gravitational constant, while the $Re$ subscript defines the effective radius aperture within which $\lambda_{Re}$ is measured. Spin parameter has become popular observationally for quantifying angular momentum, particularly in early-type galaxies \citep[as in][and many other integral field spectroscopic works since]{Emsellem+2007,Burkert+2016}. We note that it is straightforward to calculate $\lambda$ of the halo, but estimating the spin parameter of the observed galaxy requires further assumptions to derive total mass and energy of the system, including the coupling of the disk and halo spins \citep{Scannapieco+2009,Sharma+2012,Teklu+2015,Cortese+2016}. Indeed, the global theoretical value has low dynamic range \citep[0.22 dex;][]{Bullock+2001} around $\lambda \sim 0.05$, and has no strong correlation with halo mass \citep{Barnes+1987} or morphology \citep[traced by concentration parameter;][]{Bullock+2001}, though some spin-morphology trend is seen observationally, particularly for low-S{\'e}rsic-index galaxies \citep{Cortese+2016,FR18}.  \citet{Falcon-Barroso+2019} showed that aperture kinematic spin parameter $\lambda_{Re}$ does not monotonically correlate with photometric morphology quantified by $n_b$ or $\beta$. Specifically, early-type galaxies in CALIFA have $\lambda_{Re}$ that are consistent with the literature, and high-mass, highly star-forming Sb galaxies have higher $\lambda_{Re}$ as expected, but low-mass Sc and Sd galaxies have lower $\lambda_{Re}$. 
Moreover, the spin profile $\lambda(R)$ is less sensitive to local features in the halo outskirts than the ratio of rotational velocity to dispersion support $V(R)/\sigma(R)$ \citep{Wu+2014}. }
Recently, \citet{Zhu+2018NatAs,Zhu+2018MNRAS} used orbit-superposition Schwarzschild models to fit stellar orbits of CALIFA galaxies and decomposed these into a continuum of cold, warm, hot and counter-rotating kinematic components based on the PDF of circularity $\lambda_z$. They found no correlation between S{\'e}rsic index and bulge type (as traced by the kinematically hot component), contrary to \citet{fisher2008,fisher2016} who found S{\'e}rsic index to be the best predictor of bulge type. 
We note that although there are similarities between \citet{Zhu+2018MNRAS} and this work, in that both assess the relation between a galaxy's morphology and the PDF of its kinematic properties, our work uses measured specific angular momentum $j_*$ rather than modelled circularity $\lambda_z$ and does not require fitting a functional form. {Thus \pdf\/ may be preferred for morphological tracing.}

A near-universal halo angular momentum distribution was predicted by \citet{vandenBosch+2001,Bullock+2001,Liao+2017} { which is consistent with linear tidal-torque theory or angular momentum transport from minor mergers.}
However, we have seen above that \pdf\/ is not universal for the baryons in general, whereby $b_1$ correlates with $T$ and $\beta$. One could consider that dispersion-supported (classical) bulges may be more likely to approximate the angular momentum distribution of dispersion-supported DM haloes than galaxies as a whole do, but a similar non-universality is seen for bulge components, in that the skewness of the radius-selected bulge \pdf\/ $b_{1rb}$ traces $T$, $\beta$ and bulge S{\'e}rsic index $n_b$.
We note that all of the galaxies known to host pseudobulges have $b_{1rb}<0.62$, and suggest that this line might represent a clear distinction in bulge type (that is, all pseudobulges with $b_{1rb}<0.62$ and all classical bulges $b_{1rb}>0.62$. Such a distinction would restrict classical bulges to the top-right quadrant of Figure~\ref{fig:bulgen_kurtrb}, restricting the spread in $b_{1rb}$ and thus indicating a more universal \pdf\/ for classical bulges. 


A galaxy's \pdf\/ is a byproduct of its evolution -- the tidal torques experienced by the baryons and their dark matter halo, as well as the physical processes that affect the baryons after decoupling from the halo. At the peak of cosmic star formation, during $1<z<3$, stable, albeit clumpy, disks are starting to settle amongst the frequent mergers and irregular systems  \citep{Glazebrook+1995b,Driver+1995,Abraham+1996b,Abraham+1996a,Conselice+2000,Elmegreen+2005,Yuan+2017}. The disks tend to have similar-to-low specific angular momentum for their mass \citep[][]{Obreschkow+2015,Swinbank+2017,Sweet+2019,Gillman+2019}, and have high gas surface densities \citep{Daddi+2010,Tacconi+2013,Burkert+2016}, fuelling high rates of star-formation \citep{Bell+2005,Juneau+2005,Swinbank+2009,Genzel+2011} and enhanced turbulence \citep{Forster-Schreiber+2009,Wisnioski+2011,Wuyts+2012,Fisher+2014} within giant, star-forming clumps \citep{Obreschkow+2016,Romeo+2018,Behrendt+2019}. These self-rotating clumps may remain bound while migrating towards the galaxy centre  \citep{Ceverino+2012}, contributing to growth of the (pseudo)bulge and moving the galaxy towards the $z\sim0$ $\beta-j_*/M_*$ `pseudobulge track' \citep{Sweet+2019}.
Since the \pdf\/ encodes the evolutionary history of the galaxy, we suggest that it may be used to single out galaxy components and witness these key evolutionary stages. 
In a forthcoming paper (Espejo et al., in prep) we will illustrate that using circular-only velocity at high-redshift results in a \pdf\/ that is sensitive to a thin disk-like component, even in the presence of giant, star-forming clumps. This technique could thus trace the emergence of disks predicted at this epoch. Further, we have demonstrated a proof-of-concept {separation of the bulge and disk \pdf\/}, where we used a photometric $\beta$ prior to select `bulge' material in a radius-based method. Future work will develop these methods with an iterative approach to find a photokinematic $\beta$, which may then yield a powerful, physically-grounded method for decomposing galaxies in a photokinematic manner, separating clumps, classical bulges and pseudobulges, thin and thick disks.

\section{Summary and Conclusions}
\label{sec:conclusions} 

In this work we have shown, using a sample of normal local galaxies from the CALIFA survey, that a galaxy's internal distribution of spatially-resolved stellar specific angular momentum, \pdf\/, is a tracer of its photometric morphology. 

Our findings are summarised as follows: \begin{enumerate}
    \item We confirm previous results that \pdf\/ is not well described by the corresponding angular momentum distribution of a dark matter halo in a $\Lambda$CDM universe, for any morphological type, even though $j_*$ is empirically and coincidentally similar to $j_h$, depending on the assumed functional form of the stellar mass -- halo mass relation \citep{Barnes+1987,Catelan+1996a,Catelan+1996b,vandenBosch+2001,Posti2018MN}. The mismatch between stellar \pdf\/ and halo \pdfh\/ may be explained by physical processes that remove high- or low-angular momentum material and transport angular momentum throughout the disk, illustrating that galaxy assembly is dominated by such processes. \pdf\/ is also not well described by an analytic solution for an exponential disk with uniform rotational velocity, or by a normal + log-normal distribution fitting function. 
    \item The shape of \pdf\/ is characterised by skewness $b_1$, which advantageously does not depend on bin size since it is calculated on the underlying distribution, and does not require fitting as it is non-parametric.
    \item \pdf\/ is not universal. Rather, $b_1$ is positively correlated with Hubble type $T$ and bulge-to-total light ratio $\beta$ with the following best-fitting relations: \\
    $b_1 = 0.82(0.10) + 0.06(0.02)T$, with scatter $\sigma = 0.20(0.03)$; \\
    $b_1 = 0.81(0.09) + 1.52(0.32)\beta$, with $\sigma = 0.09(0.02)$; \\the quantities in parentheses represent 1-$\sigma$ uncertainties. That is, earlier-type, more bulge-dominated galaxies have more strongly-skewed \pdf\/.
    \item We decompose the \pdf\/ into `bulge' and `disk' components using a photometric prior for $\beta$ and the radius-selection method.
    \item $b_{1rb}$ is correlated with Hubble type $T$ and bulge-to-total light ratio $\beta$ with the following best-fitting relations: \\
    $b_{1rb}~=~0.43(0.10) + 0.03(0.02)T$, with scatter $\sigma = 0.17(0.03)$; \\
    $b_{1rb}~=~0.35(0.05) + 1.03(0.17)\beta$, with $\sigma = 0.05(0.02)$. 
    \item Bulge \pdf\/ skewness $b_{1rb}$ correlates with bulge S{\'e}rsic index $n_b$, with $b_{1rb} = 0.34(0.06) + 0.13(0.03)n_b$ and $\sigma = 0.12(0.02)$. Almost all galaxies that are known to host pseudobulges ($n_b <2.2$) have bulge \pdf\/ $b_{1rb}\leq 0.62$.
    \item In the absence of reliable dispersion measurements (e.g. at high redshift\footnote{It is not yet known whether or not the relation between $b_{1c}$ and $b_1$ holds for high-redshift galaxies, which are more dispersion-dominated and whose observations are more affected by beam-smearing.}), the circular-velocity \pdf\/ can be calculated. Circular \pdf\/ skewness $b_{1c}$ underestimates dispersion-broadened skewness $b_1$, but the tight correlation (coefficient $R=0.94$; scatter $\sigma = 0.07 \pm 0.01$) allows reliable conversion using the relation $b_{1c}~=~-0.82(0.13)~+~1.51(0.11)b_1$.
\end{enumerate}

The \pdf\/ traces photokinematic morphology of a galaxy. It encodes more physical information than photometry alone, and requires fewer assumptions than spin parameter $\lambda$. At high redshifts, due to decreased angular size and surface brightness dimming, photometric morphology is even more dependent on who is doing the classification, and singling out a thin disk is even more difficult than for local galaxies. \pdf\/ is a product of the evolutionary history of the baryons in the galaxy, and thus provides a classifier-independent, physically-grounded tracer of galaxy morphology. 

In forthcoming papers we will explore a sample of $z\sim1.5$ galaxies in COSMOS (Espejo et al.) and further investigate the fundamental link between \pdf\/ and morphology (Sweet et al.). As it matures, \pdf\/ will become a useful method to separate out kinematic components: thin disk from thick disk and bulge, clumps from bulges, and pseudobulges from classical bulges, facilitating observation of the emergence of the thin disk at high redshift and tracing the morphological evolution of clumps to today's bulges.

\section*{Acknowledgements}
SMS, KG and CL have received funding from the ARC Centre of
Excellence for All Sky Astrophysics in 3 Dimensions (ASTRO 3D), through project number CE170100013.  
SMS, KG, DO and DBF acknowledge support from ARC DP grant DP160102235.
DBF acknowledges support from ARC  Future  Fellowship FT170100376.
AB thanks the colleagues at Swinburne University of Technology for their hospitality and support during his visit.

This study uses data provided by the Calar Alto Legacy Integral Field Area (CALIFA) survey (http://califa.caha.es/). Based on observations collected at the Centro Astron{\'o}mico Hispano Alem{\'a}n (CAHA) at Calar Alto, operated jointly by the Max-Planck-Institut f{\"u}r Astronomie and the Instituto de Astrof{\'i}sica de Andaluc{\'i}a (CSIC).

Funding for the Sloan Digital Sky Survey (SDSS) has been provided by the Alfred P. Sloan Foundation, the Participating Institutions, the National Aeronautics and Space Administration, the National Science Foundation, the U.S. Department of Energy, the Japanese Monbukagakusho, and the Max Planck Society. The SDSS Web site is http://www.sdss.org/.

The SDSS is managed by the Astrophysical Research Consortium (ARC) for the Participating Institutions. The Participating Institutions are The University of Chicago, Fermilab, the Institute for Advanced Study, the Japan Participation Group, The Johns Hopkins University, Los Alamos National Laboratory, the Max-Planck-Institute for Astronomy (MPIA), the Max-Planck-Institute for Astrophysics (MPA), New Mexico State University, University of Pittsburgh, Princeton University, the United States Naval Observatory, and the University of Washington.




\bibliographystyle{mnras}
\bibliography{references} 



\appendix

\section{SDSS images}
\label{app:thumbs}

\begin{figure}
\begin{center}
\includegraphics[width=0.45\linewidth]{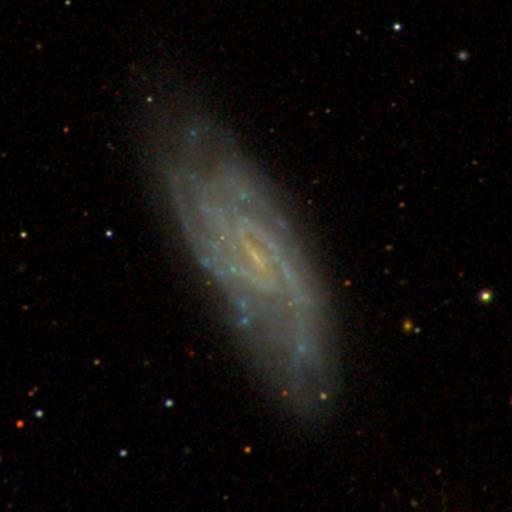}
\caption{IC1151. $T_{\rm Fisher} = $ Scd; $T_{\rm RC3} = $ Scd. 
}\label{fig:IC1151}
\end{center}
\end{figure}

\begin{figure}
\begin{center}
\includegraphics[width=0.45\linewidth]{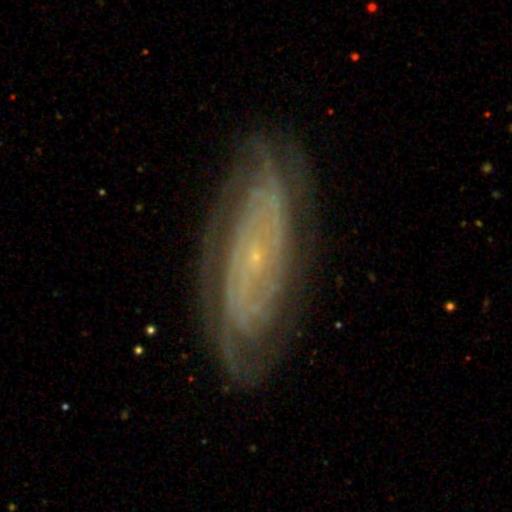}
\caption{MCG-02-02-030. $T_{\rm Fisher} = $ Sb; $T_{\rm RC3} = $ Sb. 
}\label{fig:MCG-02-02-030}
\end{center}
\end{figure}

\begin{figure}
\begin{center}
\includegraphics[width=0.45\linewidth]{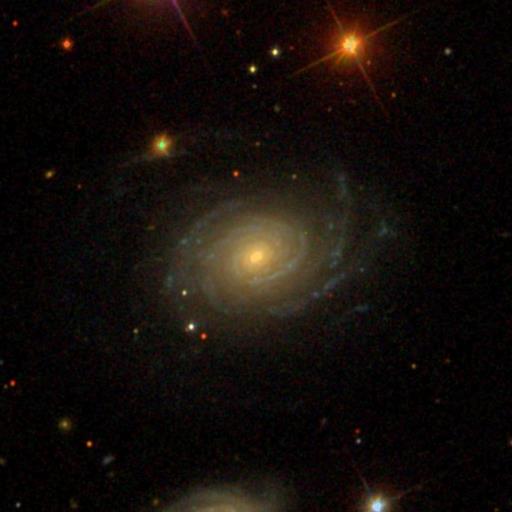}
\caption{NGC0001. $T_{\rm Fisher} = $ Sb; $T_{\rm RC3} = $ Sb. 
}\label{fig:NGC0001}
\end{center}
\end{figure}

\begin{figure}
\begin{center}
\includegraphics[width=0.45\linewidth]{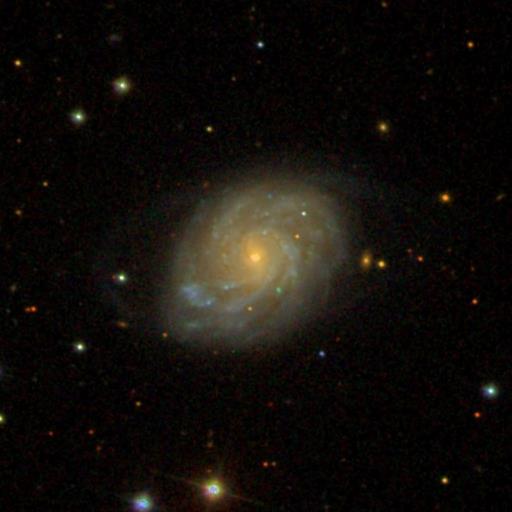}
\caption{NGC2253. $T_{\rm Fisher} = $ Sbc; $T_{\rm RC3} = $ Sc. 
}\label{fig:NGC2253}
\end{center}
\end{figure}

\begin{figure}
\begin{center}
\includegraphics[width=0.45\linewidth]{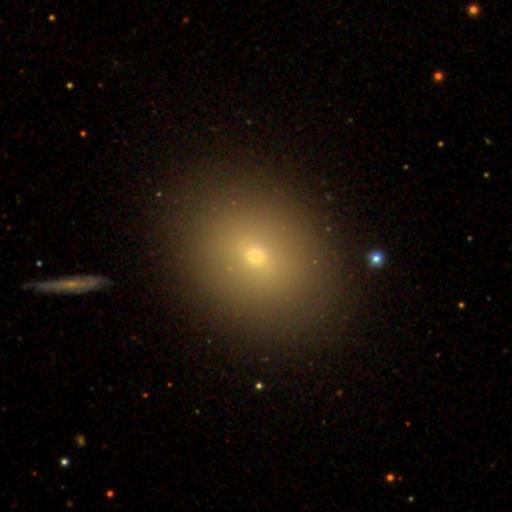}
\caption{NGC2592. $T_{\rm Fisher} = $ E?S0?; $T_{\rm RC3} = $ E. 
}\label{fig:NGC2592}
\end{center}
\end{figure}

\begin{figure}
\begin{center}
\includegraphics[width=0.45\linewidth]{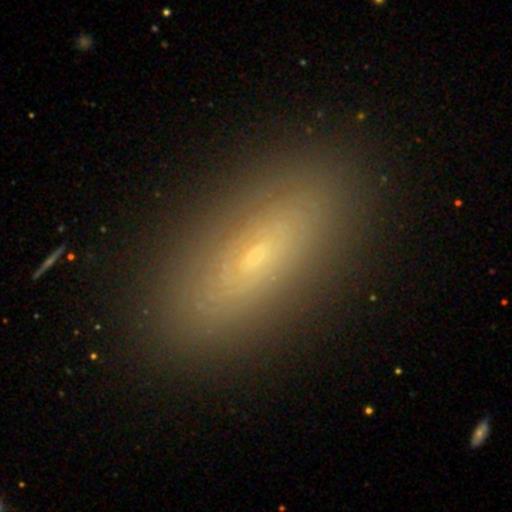}
\caption{NGC2639. $T_{\rm Fisher} = $ Sa; $T_{\rm RC3} = $ Sa. 
}\label{fig:NGC2639}
\end{center}
\end{figure}

\begin{figure}
\begin{center}
\includegraphics[width=0.45\linewidth]{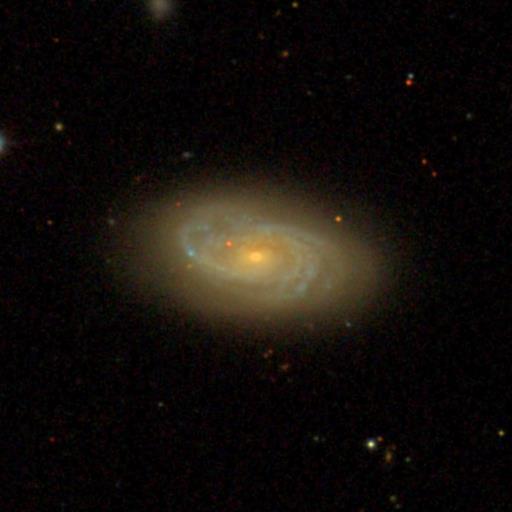}
\caption{NGC2906. $T_{\rm Fisher} = $ Sbc; $T_{\rm RC3} = $ Sc. 
}\label{fig:NGC2906}
\end{center}
\end{figure}

\begin{figure}
\begin{center}
\includegraphics[width=0.45\linewidth]{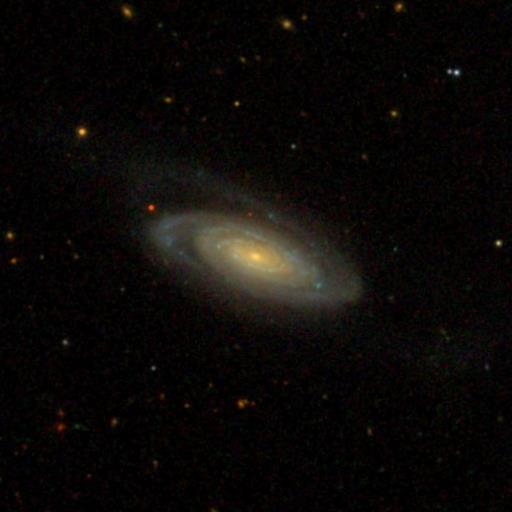}
\caption{NGC3815. $T_{\rm Fisher} = $ Sab; $T_{\rm RC3} = $ Sab. 
}\label{fig:NGC3815}
\end{center}
\end{figure}

\begin{figure}
\begin{center}
\includegraphics[width=0.45\linewidth]{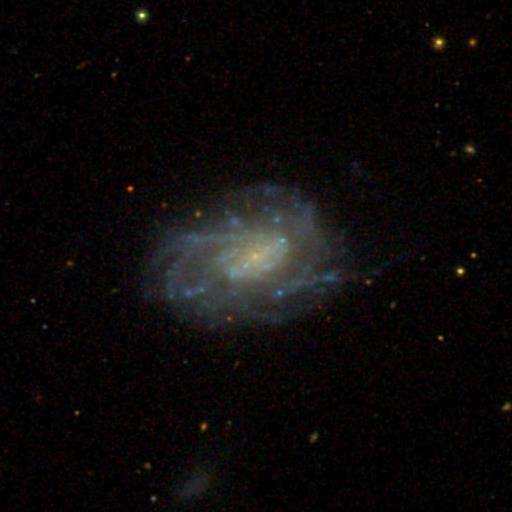}
\caption{NGC4961. $T_{\rm Fisher} = $ Sc; $T_{\rm RC3} = $ Sc. 
}\label{fig:NGC4961}
\end{center}
\end{figure}

\begin{figure}
\begin{center}
\includegraphics[width=0.45\linewidth]{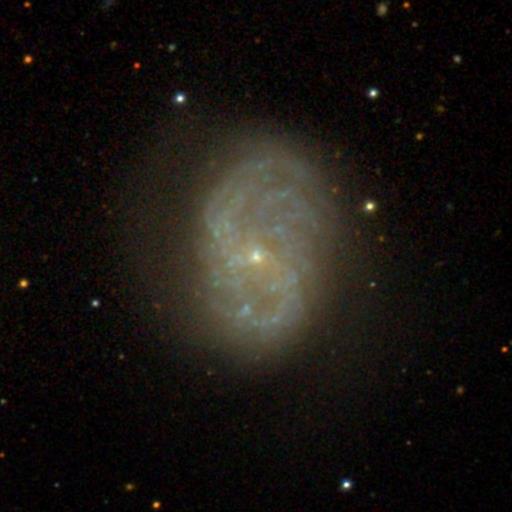}
\caption{NGC5480. $T_{\rm Fisher} = $ Sd; $T_{\rm RC3} = $ Sc. 
}\label{fig:NGC5480}
\end{center}
\end{figure}

\begin{figure}
\begin{center}
\includegraphics[width=0.45\linewidth]{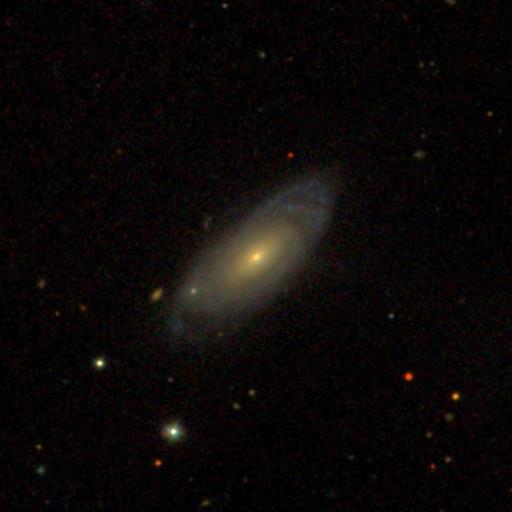}
\caption{NGC5971. $T_{\rm Fisher} = $ Sa; $T_{\rm RC3} = $ Sa. 
}\label{fig:NGC5971}
\end{center}
\end{figure}

\begin{figure}
\begin{center}
\includegraphics[width=0.45\linewidth]{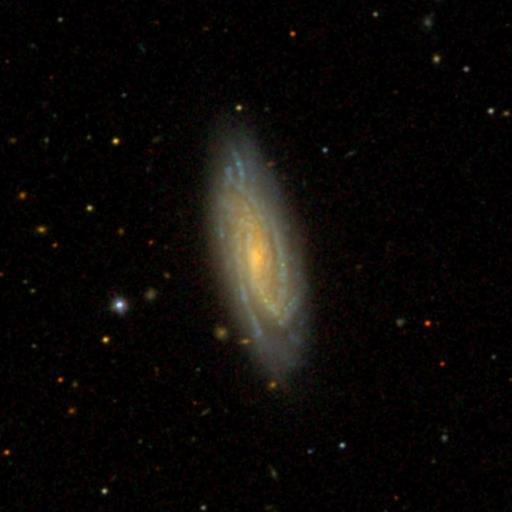}
\caption{NGC5980. $T_{\rm Fisher} = $ Sbc; $T_{\rm RC3} = $ Sbc. 
}\label{fig:NGC5980}
\end{center}
\end{figure}

\begin{figure}
\begin{center}
\includegraphics[width=0.45\linewidth]{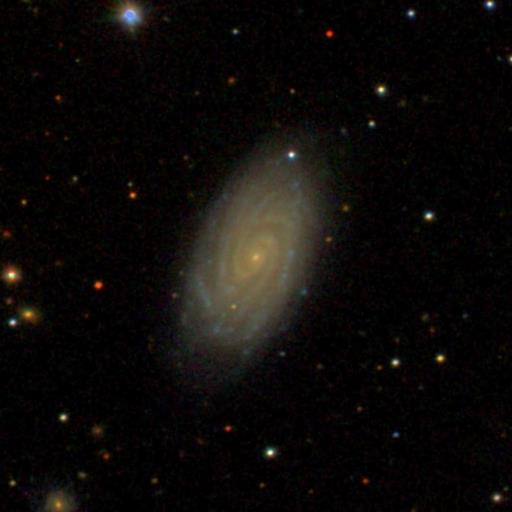}
\caption{NGC6063. $T_{\rm Fisher} = $ Sc; $T_{\rm RC3} = $ Sc. 
}\label{fig:NGC6063}
\end{center}
\end{figure}

\begin{figure}
\begin{center}
\includegraphics[width=0.45\linewidth]{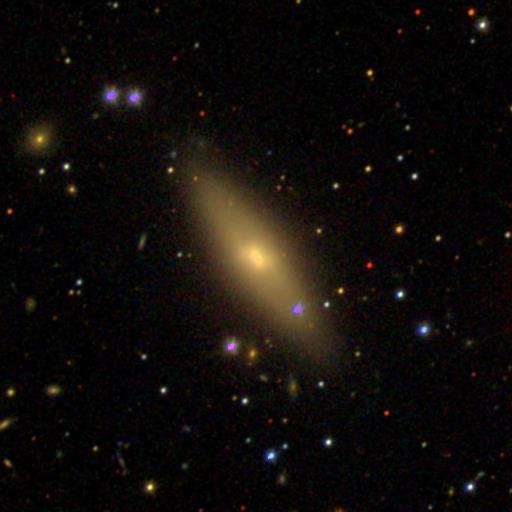}
\caption{NGC6427. $T_{\rm Fisher} = $ S0; $T_{\rm RC3} = $ E-S0. 
}\label{fig:NGC6427}
\end{center}
\end{figure}

\begin{figure}
\begin{center}
\includegraphics[width=0.45\linewidth]{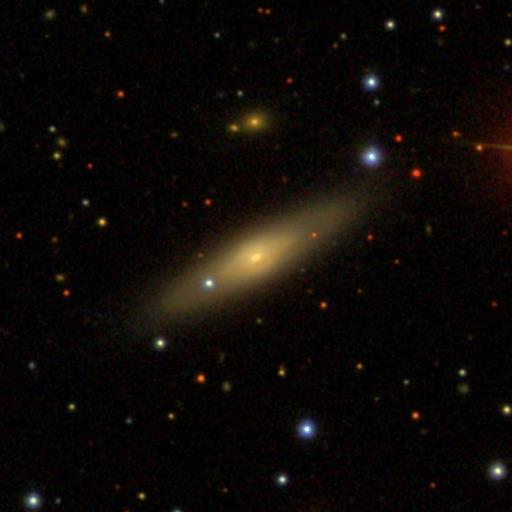}
\caption{NGC6762. $T_{\rm Fisher} = $ S0; $T_{\rm RC3} = $ S0-a. 
}\label{fig:NGC6762}
\end{center}
\end{figure}

\begin{figure}
\begin{center}
\includegraphics[width=0.45\linewidth]{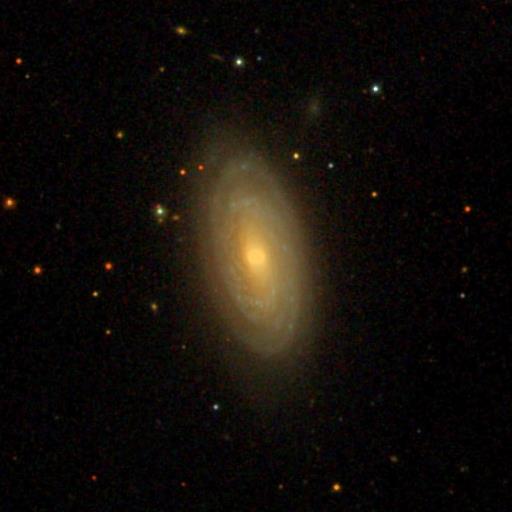}
\caption{NGC7311. $T_{\rm Fisher} = $ Sab; $T_{\rm RC3} = $ Sab. 
}\label{fig:NGC7311}
\end{center}
\end{figure}

\begin{figure}
\begin{center}
\includegraphics[width=0.45\linewidth]{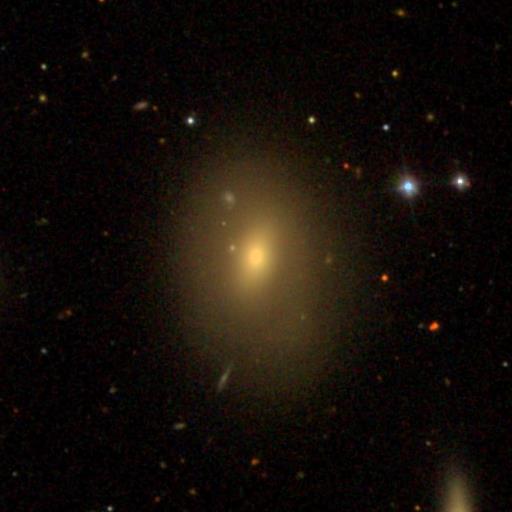}
\caption{NGC7623. $T_{\rm Fisher} = $ S0; $T_{\rm RC3} = $ S0. 
}\label{fig:NGC7623}
\end{center}
\end{figure}

\begin{figure}
\begin{center}
\includegraphics[width=0.45\linewidth]{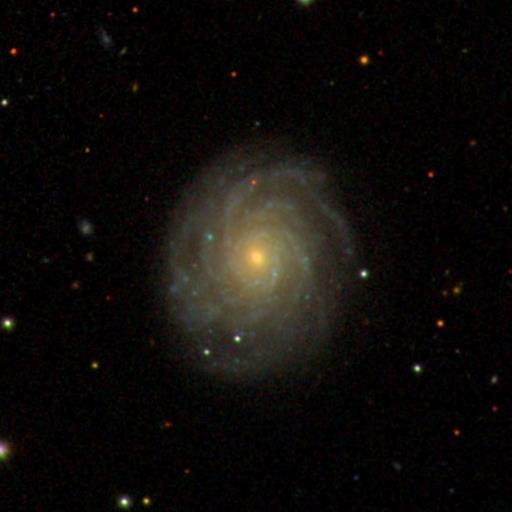}
\caption{NGC7653. $T_{\rm Fisher} = $ Sb; $T_{\rm RC3} = $ Sb. 
}\label{fig:NGC7653}
\end{center}
\end{figure}

\begin{figure}
\begin{center}
\includegraphics[width=0.45\linewidth]{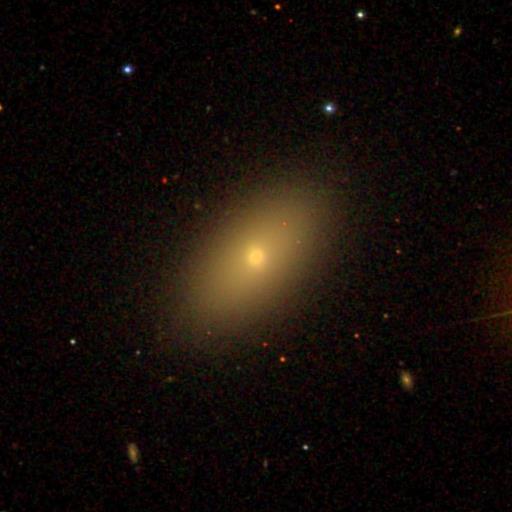}
\caption{NGC7671. $T_{\rm Fisher} = $ S0; $T_{\rm RC3} = $ S0. 
}\label{fig:NGC7671}
\end{center}
\end{figure}

\begin{figure}
\begin{center}
\includegraphics[width=0.45\linewidth]{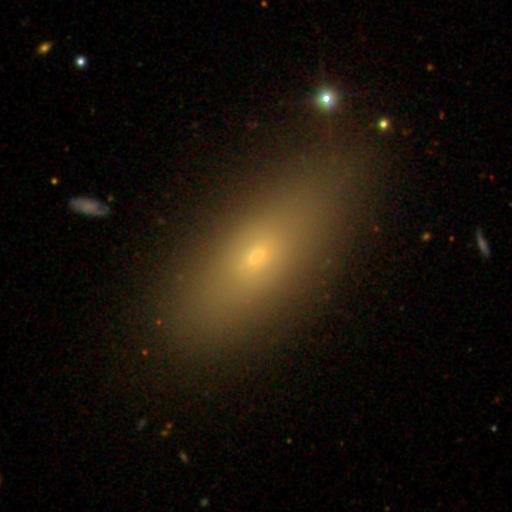}
\caption{NGC7683. $T_{\rm Fisher} = $ S0; $T_{\rm RC3} = $ S0. 
}\label{fig:NGC7683}
\end{center}
\end{figure}

\begin{figure}
\begin{center}
\includegraphics[width=0.45\linewidth]{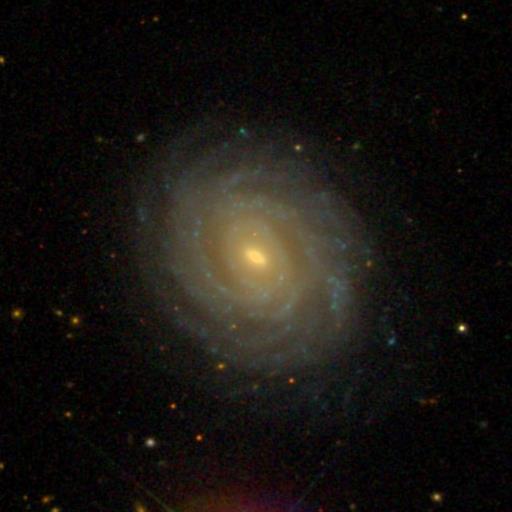}
\caption{NGC7716. $T_{\rm Fisher} = $ Sbc; $T_{\rm RC3} = $ Sb. 
}\label{fig:NGC7716}
\end{center}
\end{figure}

\begin{figure}
\begin{center}
\includegraphics[width=0.45\linewidth]{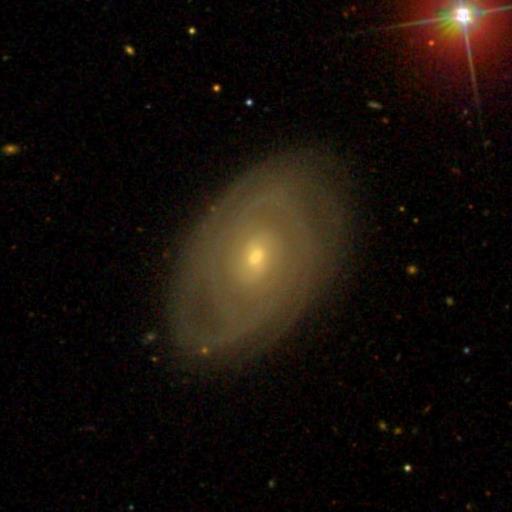}
\caption{NGC7824. $T_{\rm Fisher} = $ Sab; $T_{\rm RC3} = $ Sab. 
}\label{fig:NGC7824}
\end{center}
\end{figure}

\begin{figure}
\begin{center}
\includegraphics[width=0.45\linewidth]{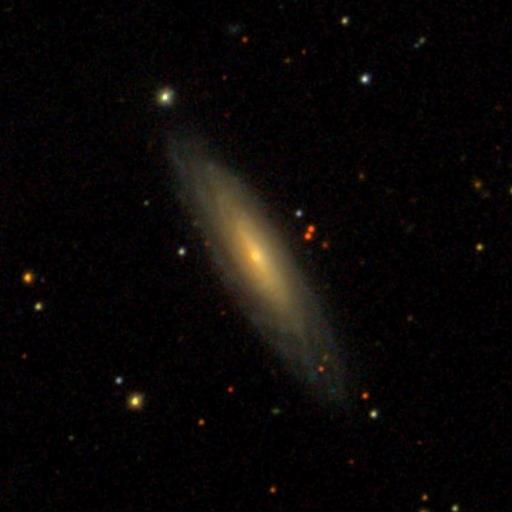}
\caption{UGC00987. $T_{\rm Fisher} = $ Sa; $T_{\rm RC3} = $ Sa. 
}\label{fig:UGC00987}
\end{center}
\end{figure}

\begin{figure}
\begin{center}
\includegraphics[width=0.45\linewidth]{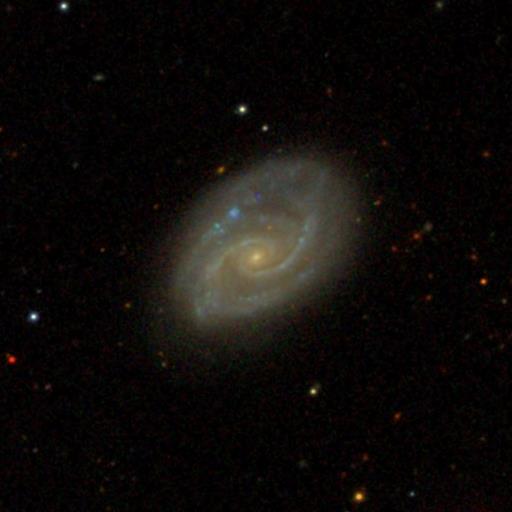}
\caption{UGC09476. $T_{\rm Fisher} = $ Sc; $T_{\rm RC3} = $ SABc. 
}\label{fig:UGC09476}
\end{center}
\end{figure}

We show an SDSS image cutout of each of the galaxies in our sample in Figures~\ref{fig:IC1151} to~\ref{fig:UGC09476}. Each caption contains Hubble type $T_{\rm Fisher}$ from our method, as well as $T_{\rm RC3}$ from the Third Reference Catalogue of Bright Galaxies \citep[RC3,][]{RC3} for comparison. Each image is 205$\arcsec$ on a side. N is up; E is left.

\bsp    
\label{lastpage}
\end{document}